\documentclass[aps,twocolumn,showpacs]{revtex4}

\usepackage{bm}
\usepackage{epsfig}
\usepackage[psamsfonts]{amsfonts,amssymb}
\usepackage[psamsfonts,mathscr]{eucal}
\newcommand{\Def}{\newcommand}
\Def{\DeF}{\renewcommand}
\Def{\Thm}{\newtheorem}

\Def{\bq}{\begin{equation}}
\Def{\eq}{\end{equation}}
\Def{\bQ}{\begin{eqnarray}}
\Def{\eQ}{\end{eqnarray}}

\Thm{definition}{Definition}
\Thm{application}{Application}
\Thm{hypothesis}{Hypothesis}
\Thm{lemma}{Lemma}
\Thm{problem}{Problem}
\Thm{proposition}{Proposition}
\Thm{remark}{Remark}
\Thm{theorem}{Theorem}

\DeF{\AA}{\mathscr{A}}
\Def{\BB}{\mathscr{B}}
\Def{\CC}{\mathscr{C}}
\Def{\DD}{\mathscr{D}}
\Def{\EE}{\mathscr{E}}
\Def{\FF}{\mathscr{F}}
\Def{\GG}{\mathcal{G}}
\Def{\HH}{\mathscr{H}}
\Def{\II}{\mathscr{I}}
\Def{\JJ}{\mathscr{J}}
\Def{\KK}{\mathscr{K}}
\Def{\LL}{\mathscr{L}}
\Def{\MM}{\mathscr{M}}
\Def{\NN}{\mathscr{N}}
\Def{\OO}{\mathscr{O}}
\Def{\PP}{\mathscr{P}}
\Def{\QQ}{\mathscr{Q}}
\Def{\RR}{\mathscr{R}}
\DeF{\SS}{\mathscr{S}}
\Def{\TT}{\mathscr{T}}
\Def{\UU}{\mathscr{U}}
\Def{\VV}{\mathscr{V}}
\Def{\WW}{\mathscr{W}}
\Def{\XX}{\mathscr{X}}
\Def{\YY}{\mathscr{Y}}
\Def{\ZZ}{\mathscr{Z}}

\Def{\bbA}{\mathbb{A}}
\Def{\bbB}{\mathbb{B}}
\Def{\bbC}{\mathbb{C}}
\Def{\bbD}{\mathbb{D}}
\Def{\bbE}{\mathbb{E}}
\Def{\bbF}{\mathbb{F}}
\Def{\bbG}{\mathbb{G}}
\Def{\bbH}{\mathbb{H}}
\Def{\bbI}{\mathbb{I}}
\Def{\bbJ}{\mathbb{J}}
\Def{\bbK}{\mathbb{K}}
\Def{\bbL}{\mathbb{L}}
\Def{\bbM}{\mathbb{M}}
\Def{\bbN}{\mathbb{N}}
\Def{\bbO}{\mathbb{O}}
\Def{\bbP}{\mathbb{P}}
\Def{\bbQ}{\mathbb{Q}}
\Def{\bbR}{\mathbb{R}}
\Def{\bbS}{\mathbb{S}}
\Def{\bbT}{\mathbb{T}}
\Def{\bbU}{\mathbb{U}}
\Def{\bbV}{\mathbb{V}}
\Def{\bbW}{\mathbb{W}}
\Def{\bbX}{\mathbb{X}}
\Def{\bbY}{\mathbb{Y}}
\Def{\bbZ}{\mathbb{Z}}

\Def{\And}{\mathrm{and}}
\Def{\Der}{\mathrm{Der}}
\Def{\ID}{\mathbf{1}}
\Def{\ie}{\mathrm{i.e.\,}}
\Def{\If}{\mathrm{If}}
\Def{\ii}{\mathbf{i}}
\Def{\implies}{\Longrightarrow}
\Def{\Ker}{\mathrm{Ker\,}}
\Def{\non}{\nonumber}
\Def{\QED}{\emph{Q.E.D.} \mbox{\rule[-1.5pt]{6pt}{10pt}}}
\Def{\Rg}{\mathrm{Rg\,}}
\Def{\st}{\mathrm{s.t.\,}}
\Def{\Then}{\mathrm{Then}}
\Def{\tr}{\mathrm{tr\,}}
\Def{\where}{\mathrm{where}}

\DeF{\theequation}{\thesection.\arabic{equation}}
\DeF\refname{}

\begin{document}
\title{Control of Hamiltonian chaos as a possible tool \\
to control anomalous transport in fusion plasmas}
\author{Guido Ciraolo}\email{ciraolo@arcetri.astro.it}
\affiliation{Facolt\`a di Ingegneria, Universit\`a di Firenze,
via S. Marta, I-50129 Firenze, Italy, and I.N.F.M. UdR Firenze}
\author{Fran\c{c}oise Briolle}\email{briolle@cpt.univ-mrs.fr}
\affiliation{CPT-CNRS, Luminy Case 907, F-13288 Marseille Cedex 9,
France}
\author{Cristel Chandre}\email{chandre@cpt.univ-mrs.fr}
\affiliation{CPT-CNRS, Luminy Case 907, F-13288 Marseille Cedex 9,
France}
\author{Elena Floriani}\email{floriani@cpt.univ-mrs.fr}
\affiliation{CPT-CNRS, Luminy Case 907, F-13288 Marseille Cedex 9,
France}
\author{Ricardo Lima}\email{lima@cpt.univ-mrs.fr}
\affiliation{CPT-CNRS, Luminy Case 907, F-13288 Marseille Cedex 9,
France}
\author{Michel Vittot}\email{vittot@cpt.univ-mrs.fr}
\affiliation{CPT-CNRS, Luminy Case 907, F-13288 Marseille Cedex 9,
France}
\author{Marco Pettini}\email{pettini@arcetri.astro.it}
\affiliation{Istituto Nazionale di Astrofisica, Osservatorio
Astrofisico di Arcetri, Largo Enrico Fermi 5, I-50125 Firenze, Italy,
I.N.F.M. UdR Firenze and I.N.F.N. Sezione di Firenze}
\author{Charles Figarella}\email{charles.figarella@cea.fr}
\affiliation{Association Euratom-CEA, DRFC/DSM/CEA, CEA Cadarache,
F-13108 St. Paul-lez-Durance Cedex, France}
\author{Philippe Ghendrih}\email{philippe.ghendrih@cea.fr}
\affiliation{Association Euratom-CEA, DRFC/DSM/CEA, CEA Cadarache,
F-13108 St. Paul-lez-Durance Cedex, France}
\date{\today}
\begin{abstract}
It is shown that a relevant control of Hamiltonian chaos is possible through
suitable small perturbations whose form can be explicitly computed. In
particular, it is possible to control (reduce) the
chaotic diffusion in the phase space of a Hamiltonian system with
$1.5$ degrees of freedom which models the diffusion of charged 
test particles in a
turbulent electric field across the confining magnetic field in
controlled thermonuclear fusion devices. Though still far from
practical applications, this result suggests that some strategy to
control turbulent transport in magnetized plasmas, in particular
tokamaks, is conceivable.
The robustness of the control is investigated in terms of a departure from
the optimum magnitude, of a varying cut-off at large wave vectors, and of
random errors on the phases of the modes. In all three cases, there is a
significant region of maximum efficiency in the vicinity of the optimum
control term.
\end{abstract}

\maketitle

\section{Introduction}
Transport induced by chaotic motion is now a
standard framework to analyze the properties of numerous systems.
Since chaos  can be harmful in several contexts, during the last decade 
or so, much attention has been paid to the so-called topic of
{\it chaos control}.
Here the meaning of {\em control} is that one aims
at reducing or suppressing chaos inducing a relevant change
in the transport properties, by means of a small perturbation 
(either open-loop 
or closed-loop control of dissipative systems \cite{limapet,review}) 
so that the original structure of
the system under investigation is substantially kept unaltered.
Control of {\em chaotic transport} properties still remains an open issue
with considerable applications.\\ 
\indent In the case of dissipative systems, 
an efficient strategy of control works by stabilizing unstable 
periodic orbits, 
where the dynamics is eventually attracted. Similarly,
a first idea to control Hamiltonian systems is to modify the parameters of the system in order to act on periodic orbits: One can enhance the stability of elliptic periodic orbits by zeroing their residues~\cite{caha}, or by stabilizing
 hyperbolic periodic orbits~\cite{upo}. \\
Another idea to stabilize the system is to enlarge the phase space by coupling the system with an external system (and hence with additional degrees of freedom
which makes the large system more regular)~\cite{emb}. 
These embedding techniques are similar to the above methods on the 
stabilization of unstable periodic orbits; they are based on the 
construction of a dissipative system from the original Hamiltonian system. 
The techniques developed for dissipative systems 
can thus be applied to this modified system, like for instance the targeting 
of periodic orbits.\\
A different approach is to modify the Hamiltonian (or just the potential) to control the original system. This approach is useful when one is able to act on this system with an external forcing. The interesting point is that the Hamiltonian structure with its number of degrees of freedom is preserved. So far, the modifications of the Hamiltonian that have been proposed in the literature are: the modification of the integrable part of the Hamiltonian~\cite{gallavotti}, the control of a system with large and non-smooth external pulses~\cite{modPulse}, a localised control with a modification in some specific regions of phase space~\cite{modLoc}, or a control using variations of the external field~\cite{opt,modExt}. However, we notice that most of the modifications of the potential that have been proposed so far are tailored to specific examples (with the exception of the optimal control~\cite{opt}) and the required modifications are large compared with the potential.\\
Hamiltonian description of the microscopic origin of particle transport usually involves
a large number of particles. Methods based on
targeting and locking to islands of regular motions in a 
``chaotic sea'' are of no practical use in control when
dealing simultaneously with a large number of unknown trajectories.
Therefore, the most efficient procedure appears to be to control the transport
process with a small perturbation, if any,
making the system integrable or closer to integrable.
In what follows we show that it is actually possible
to control Hamiltonian chaos in this way by preserving the Hamiltonian 
structure. We describe a general method for controlling 
nearly integrable Hamiltonian 
systems, and we apply this tecnique to a model relevant to 
magnetized plasmas.
\\ \indent Chaotic transport  of particles advected by a turbulent
electric field with a strong magnetic field
is associated with Hamiltonian dynamical systems 
under the approximation of the guiding center motion due to 
${\bf E \times B}$ drift velocity.
For an appropriate choice of the turbulent electric field, it has
been shown that the resulting diffusive transport is then found to agree
with the experimental counterpart \cite{marc88}.
It is clear that such an
analysis is only a first step in the investigation and
understanding of turbulent plasma transport.
The control of
transport in magnetically confined plasmas is of major importance
in the long way to achieve 
controlled thermonuclear fusion. Two major mechanisms
have been proposed for such a turbulent transport: transport
governed by the fluctuations of the magnetic field and transport
governed by fluctuations of the electric field. There is presently
a general  consensus to consider, at low plasma pressure, that the
latter mechanism  agrees with experimental
evidence~\cite{scott}. In the area of
transport of trace impurities, i.e. that are sufficiently
diluted so as not to modify the electric field pattern, 
the {\em $\bf E \times B$} drift motion of test particle should be the
exact transport model. Even for this very restricted case, control
of chaotic transport would be very relevant for the
thermonuclear fusion program.
The possibility of reducing and even suppressing chaos
combined with the empirically found states of improved confinement in
tokamaks,
suggest to investigate the possibility to devise
a strategy of control of chaotic transport through
some smart perturbations acting at the microscopic level of charged
particle motions.\\
\indent  As in the current literature the electric turbulent transport 
in plasmas is mainly addressed in the Eulerian (fluid) framework, 
let us first recall the difference between
Lagrangian and Eulerian descriptions of transport. We consider
the advection of a scalar quantity
$\theta ({\bf x}, t)$ describing, e.g., the concentration of a passively 
transported entity. In a given Eulerian velocity field ${\bf v}({\bf x}, t)$
the transport of $\theta ({\bf x}, t)$ is described by
\begin{equation}
\frac{\partial\theta({\bf x}, t)}{\partial t} + {\bf v({\bf x}}, t)
\cdot\nabla \theta({\bf x}, t) =D\nabla^2\theta({\bf x}, t),
\label{advection}
\end{equation}
where $D$ is a molecular diffusion coefficient. 
This equation holds for both neutral fluids and plasmas. If the field
${\bf v}({\bf x}, t)$ is given indipendently from the field 
$\theta ({\bf x}, t)$ Eq.~(\ref{advection}) is linear in ${\bf v}({\bf x}, t)$.
The complexity of the field $\theta ({\bf x}, t)$ will then depend on both
the complexity of the field ${\bf v}({\bf x}, t)$ and on the molecular 
diffusion coefficient $D$. When considering the simulation of 
Eq.~(\ref{advection}), the magnitude of $D$ will govern the mesh size to store
the field $\theta ({\bf x}, t)$. For a vanishingly small diffusion $D$,
there will be no cut-off of the small scales generated by the simulation.
This will require an infinite storage capability to describe the complexity
of $\theta ({\bf x}, t)$ that can appear even for a relatively smooth 
velocity field. To evaluate this property, the most straightforward description
is given by a Lagrangian approach. For the same transport process, the latter
requires to solve the following
equations of motion of a passive tracer (e.g. particle, fluid drop) 
whose Eulerian concentration function is $\theta ({\bf x}, t)$,
\begin{equation}
\dot {\bf x} = {\bf v({\bf x}}, t),
\label{lagrang}
\end{equation}
which in the case of a two dimensional incompressible 
Euler flow can be given the form
\begin{equation}
{\dot{\bf x}={d\over dt}{x\choose y}={\bf v}({\bf x},t)={\nabla}^{\perp}\psi
={-\partial_y \psi (x,y,t)\choose \; \partial_x \psi (x,y,t)}},
\label{scalar}
\end{equation}
where $\psi$ denotes the stream function of the eulerian field 
${\bf v}({\bf x},t)$, the trajectory of the tracer is denoted by ${\bf x}(t)$
and $\nabla^\perp\equiv(-\partial_y, \partial_x)$. 
What is remarkable here is the Hamiltonian structure of the equations of
motion (\ref{scalar}), where the stream function $\psi$ plays the role of the
Hamiltonian function and $x$ and $y$ 
play the role of the canonically conjugate variables.
With the exception of trivial velocity fields ${\bf v}$
(like a uniform, parallel flow) these equations of motion are in general 
nonlinear in the coordinates; in fact, if we even think of a simple vortex,
we realize that $v_x$ and $v_y$ must contain at least one trigonometric 
function. Now, also without a standard (quadratic) kinetic energy term, 
this kind of Hamiltonian dynamical system displays all the rich and complex 
phenomenology of the transition between regular and chaotic motions and 
between weak and strong chaos \cite{ws_chaos}. 
Thus, even in presence of rather regular
Eulerian velocity patterns, the solutions of Eqs.~(\ref{lagrang}) and 
(\ref{scalar}) can be very complicated, with apparently no relation left with
${\bf v}({\bf x}, t)$. 
In other words, chaotic Lagrangian diffusion can take
place also in  presence of rather simple Eulerian velocity patterns.
For realistic simulation of Eq.~(\ref{advection}) a finite mesh size must 
be introduced and accordingly the diffusion coefficient $D$ must reach a finite
value to smear out the small scales that cannot be captured by the grid.
If the velocity field is characterized by a large regular structure
superimposed to small scale structures, the output of the simulation can lead
to the absence of any diffusion but the molecular one \cite{Vulpiani}.
The difficulty in the simulation of Eq.~(\ref{advection}) will then lead
to an apparent conflict with a broad experimental evidence~\cite{exp_chaos}.
The most efficient means to address the transport of passive scalars in a given
velocity field ${\bf v}({\bf x}, t)$ appears to follow a Lagrangian approach
that allows one to describe the motion at all scales in space and time.
The cost of this method will appear in the statistics that must be performed
to obtain a general property of the system whenever a single trajectory
does not allow one to capture the properties of all possible trajectories.\\
\indent When addressing plasma transport, Eulerian and Lagrangian 
approaches are combined
to provide an analysis of the transport properties. An equation similar to 
Eq.~(\ref{advection}) is coupled to a vorticity equation defining the field
${\bf v}({\bf x}, t)$~\cite{philippe1}. The Eulerian description is used to 
generate the velocity field and a Lagrangian approach is used to follow trace
impurities and trace Tritium that allows one to compare the simulations 
to experimental data~\cite{philippe2}.\\
\indent A close analogy exists between the equations
of motion of passive tracers (\ref{scalar}) and those of the guiding centers
of charged particles moving in strongly magnetized plasmas and in presence 
either of an electric field transverse to the magnetic field, or of
an inhomogeneous component of the magnetic field itself. 
The electrostatic case \cite{Northrop} is modeled by
$$
{\dot{\bf x}}={d\over dt}{x\choose y}={c\over B^2}{\bf E}({\bf x},t)\times
 {\bf B}=
{c\over B}{-\partial_y V (x,y,t)\choose \partial_x V (x,y,t)},
$$
where $V$ is the electric potential, ${\bf E}=-\nabla V$, and
${\bf B}=B {\bf e}_z$.
The magnetic case is modeled by
\begin{eqnarray}
{\dot{\bf x}}={d\over dt}{x\choose y}&&=\frac{{\bf v}_{\parallel}}{RB}
\times
\nabla \Phi_{pol} ({\bf x},t)\nonumber\\
&&=\frac{{v}_{\parallel}}{RB}
{-\partial_y \Phi_{pol} (x,y,t)\choose \partial_x \Phi_{pol} (x,y,t)},
\end{eqnarray}
where ${\bf v}_{\parallel}$ is the velocity along the field line, $R$
the major radius of the torus and $\Phi_{pol}$ the poloidal magnetic flux
divided by $2\pi$.
In both cases, the physically remarkable phenomenon -- in complete 
analogy with
the Lagrangian diffusion of passive scalars -- is that even in presence of 
rather regular space-time patterns of the electric fields or of the magnetic
inhomogeneities, the charged particles can diffuse across the magnetic field
which ceases to be confining. The dynamical instability with respect to small
variations of the initial conditions, known as deterministic chaos, is the
very source of the enhanced cross-field diffusion; 
it is ``intrinsically'' non-collisional
and it turns out to be orders of magnitude larger than the collisional one
\cite{marc88}, sometimes even many orders of magnitude larger \cite{Amato}.\\
\indent In this article, the problem we address is how to control chaotic 
diffusion in such Hamiltonian models. In some range of parameters, the 
problems can be considered as nearly integrable.  
We consider the class of Hamiltonian systems which can be written
in the form $H=H_0+\epsilon V$ that is an integrable 
Hamiltonian $H_0$ (with action-angle variables)
plus a small perturbation $\epsilon V$.
\\ \indent The problem of control
in Hamiltonian systems is the following one: For the perturbed Hamiltonian
$H_0+\epsilon V$, the aim is to devise a control term $f$ such that
the dynamics of the controlled Hamiltonian $H_0+\epsilon V+f$
has more regular trajectories (e.g.~on invariant tori) or less diffusion
than the uncontrolled one. Obviously $f=-\epsilon V$ is a solution since the resulting Hamiltonian is integrable. However, it is a useless solution
since the control is of the same magnitude of the perturbation.
For practical purposes, the desired control term should be small
(with respect to the perturbation $\epsilon V$), localized in phase space
(meaning that the subset of phase space where $f$ is non-zero is finite
or small enough),
or $f$ should be of a specific shape (e.g. a sum of given Fourier modes, 
or with a certain regularity). Moreover, the control should be as simple 
as possible in view of future implementions  in experiments.\\
\indent In Sec.~\ref{sec:2}, we explain the control theory of nearly integrable Hamiltonian systems following Ref.~\cite{michel}. We show that it is possible to construct and compute analytically a control term $f$ of order $\varepsilon^2$ such that the controlled Hamiltonian $H_c=H_0+\epsilon V+f$ is integrable. In Sec.~\ref{sec:3}, after defining the model of interest to our study in Sec.~\ref{sec:3a}, we compute analytically the first terms of the expansion of the control term in Sec.~\ref{sec:3b}. Some properties of the control term are given in Sec.~\ref{sec:3c}. A numerical study of the effect of the control term on the dynamics is done extensively in Sec.~\ref{sec:3d}. It is shown that the chosen control term is able to drastically reduce the chaotic transport. In Sec.~\ref{sec:3e}, we study the effect of some truncations that aim at  either simplifying the control term or reducing the energy input to control the system: In particular, we show that reducing the control term to its main Fourier components
 or reducing the magnitude of the control term is sufficient to 
govern a significant decrease of the chaotic transport. Thought, of course 
the optimal control is obtained with the full control term. These 
results indicates that this control of Hamiltonian 
systems is robust.
\section{Control theory of Hamiltonian systems.}
\label{sec:2}

In this section, following the framework of Ref.~\cite{michel}
we explain the control theory of Hamiltonian systems. 
Let $\mathcal A$ be the vector space
of $C^{\infty}$ real functions defined on the
phase space. 
For $H\in \mathcal A$, let $\{H\}$ be the linear operator
acting on $\mathcal A$  such that
$$
\{H\}H^{\prime}=\{H,H^{\prime}\},
$$
for any $H^{\prime}\in \mathcal A$, where $\{\cdot~,\cdot\}$ is the Poisson bracket. 
Hence $\mathcal A$ is a Lie algebra.
The time-evolution of a function $V\in {\mathcal A}$ following the flow of $H$ 
is given by
$$
\frac{dV}{dt}=\{ H\} V,
$$  
which is formally solved as
$$
V(t)=e^{t\{H\}}V(0),
$$
if $H$ is time independent, and where
$$
e^{t\{H\}}=\sum_{n=0}^{\infty}\frac{t^n}{n!}\{H\}^n.
$$
Any element $V\in\mathcal A$ 
such that \( \{{H}\}{V} =0 \), is constant under the flow of \( {H} \),
i.e.
$$
\forall t\in {\mathbb R}, \qquad e^{{t} \{{H}\}} {V} = {V}.
$$
Let us now consider a given Hamiltonian $H_0\in {\mathcal A}$.
The operator
\(\{{H_0}\} \) is not invertible since a derivation has always 
a non-trivial kernel. For
instance \( \{ {H_0} \} {H_0}^\alpha = 0 \) for any $\alpha$ such
that ${H_0}^\alpha \in \mathcal A$. 
The vector space \( \Ker\{{H_0}\} \) is the set of constants
of motion.
Hence we consider a pseudo-inverse of \( \{{H_0}\} \).
We define a linear operator $\Gamma$ on $\mathcal A$ such that
\bq
\{{H_0}\}^{2}\ \Gamma = \{{H_0}\},
\label{gamma}
\eq
i.e.
$$
\forall V\in {\mathcal A}, \qquad \{H_0,\{H_0,\Gamma V\}\}=\{H_0,V\}.
$$
The operator $\Gamma$ is not unique. Any other choice
$\Gamma^{\prime}$ satisfies that the range 
$\rm{Rg}(\Gamma^{\prime}-\Gamma)$ is included into the kernel
$\rm{Ker}(\{H_0\}^2)$.
\\ \indent We define the {\em non-resonant} operator $\mathcal N$ and the 
{\em resonant} operator $\mathcal R$ as
\begin{eqnarray*}
&& {\mathcal N} = \{H_0\}\Gamma,\\
&& {\mathcal R} = 1-{\mathcal N},
\end{eqnarray*}
where the operator $1$ is the identity in the
algebra of linear operators acting on \( \mathcal A \). 
We notice that Eq.(\ref{gamma}) becomes
$$
\{{H_0}\} \mathcal R = 0,
$$
which means that the range \( \Rg \mathcal R \) of the operator \( \mathcal R \) is
included in \( \Ker\{{H_0}\} \).
A consequence
is that ${\mathcal R} V$ is constant under the flow of $H_0$, i.e. 
$e^{t\{H_0\}}{\mathcal R}V={\mathcal R}V$. We notice that when $\{H_0\}$
and $\Gamma$ commute, $\mathcal R$ and $\mathcal N$ are projectors, i.e.
$\mathcal R^2=\mathcal R$ and $\mathcal N^2=\mathcal N$. Moreover, in this case we have $\Rg \mathcal R = \Ker\{{H_0}\} $, i.e.\ the constant of motion are the elements $\mathcal{R}V$ where $V\in \mathcal{A}$.
\\ \indent Let us now assume that $H_0$ is integrable 
with action-angle variables 
$({\bf A},\bm{\varphi})\in {\mathcal B}\times {\mathbb T}^l $ where ${\mathcal B}$ 
is an open set of $\mathbb R^l$ and ${\mathbb T}^l$ is the $l$ dimensional torus. Thus 
$H_0=H_0({\bf A})$ and the Poisson bracket $\{H,H^{\prime}\}$ between two elements $H$ and $H'$ of ${\mathcal A}$ is 
$$
\{H,H^{\prime}\}=\frac{\partial H}{\partial{\bf A}}\cdot
\frac{\partial H^{\prime}}{\partial{\bm{\varphi}}}-
\frac{\partial H}{\partial{\bm{\varphi}}}\cdot
\frac{\partial H^{\prime}}{\partial{\bf A}}.
$$
The operator $\{H_0\}$ acts on $V$ expanded as follows
$$
V=\sum_{{\bf k}\in \mathbb Z^l}V_{\bf k}({\bf A})e^{i{\bf k}\cdot{\bm\varphi}},
$$
as
$$
\{H_0\}V({\bf A},\bm{\varphi})=\sum_{\bf k}i{\bm \omega}({\bf A})\cdot{\bf k}~V_{\bf k}({\bf A})e^{i{\bf k}\cdot\bm\varphi},
$$
where 
$$
{\bm \omega}({\bf A})=\frac{\partial H_0}{\partial{\bf A}}. 
$$
A possible choice of $\Gamma$ is
\begin{equation}
\label{eqn:GV}
\Gamma V({\bf A},\bm{\varphi})=
\sum_{{\bf k}\in{\mathbb Z^l}\atop{\omega({\bf A})\cdot{\bf k}\neq0}}
\frac{V_{\bf k}({\bf A})}
{i{\bm \omega}({\bf A}) \cdot{\bf k}}~~e^{i{\bf k}\cdot{\bm\varphi}}.
\end{equation}
We notice that this choice of $\Gamma$ commutes with $\{H_0\}$.
\\ \indent For a given $V\in{\mathcal A}$, ${\mathcal R} V$ is the resonant 
part of $V$ and ${\mathcal N} V$ is the non-resonant part:
\begin{eqnarray}
&&{\mathcal R}V=\sum_{\bf k~}
V_{\bf k}({\bf A})\chi(\bm\omega({\bf A})\cdot{\bf k}=0)e^{i{\bf k}\cdot{\bm\varphi}},\label{eqn:RV}\\ 
&&{\mathcal N}V=\sum_{\bf k~}
V_{\bf k}({\bf A})\chi(\bm\omega({\bf A})\cdot{\bf k}\neq0)e^{i{\bf k}\cdot{\bm\varphi}},\label{eqn:NV}
\end{eqnarray}
where $\chi(\alpha=0)$ vanishes  when $\alpha\neq0$ and it is equal to $1$
when  $\alpha=0$.\\

From these operators defined for the integrable part $H_0$, we construct a control term for the perturbed Hamiltonian $H_0+V$ where $V\in {\mathcal A}$, i.e.
 $f$ is constructed such that $H_0+V+f$
is canonically conjugate to $H_0+\mathcal R V$.
\begin{proposition} -- ~~For $V \in {\mathcal A}$ and $\Gamma$ constructed
from $H_0$, we have the following equation
\bq
e^{\{\Gamma V\}}(H_0+V+f)=H_0+{\mathcal R} V,
\label{prop1}
\eq
where
\bq
f(V)=e^{-\{\Gamma V\}}{\mathcal R}V+\frac{1-e^{-\{\Gamma
V\}}}{\{\Gamma V\}} {\mathcal N} V -V.
\eq
\end{proposition}
We notice that the operator $(1-e^{-\{\Gamma V\}})/\{\Gamma V\}$
is well defined by the expansion
$$
\frac{1-e^{-\{\Gamma V\}}}{\{\Gamma V\}}=
\sum_{n=0}^{\infty}\frac{(-1)^n}{(n+1)!}\{\Gamma V\}^n.
$$
\\
{\em Proof}:
Since $e^{\{\Gamma V\}}$ is invertible, Eq.~(\ref{prop1}) gives
$$
f(V)=(e^{-\{\Gamma V\}}-1)H_0+e^{-\{\Gamma V\}}{\mathcal R}V-V.
$$
We notice that the operator $e^{-\{\Gamma V\}}-1$ can be divided by 
$\{\Gamma V\}$
$$
f(V)=\frac{e^{-\{\Gamma V\}}-1}{\{\Gamma V\}}\{\Gamma V\}H_0+
e^{-\{\Gamma V\}}{\mathcal R}V-V.
$$
By using the relations
$$
\{\Gamma V\}H_0=\{\Gamma V,H_0\}=-\{H_0\}\Gamma V,
$$
and
$$
\{H_0\}\Gamma={\mathcal N},
$$
we have
$$
f(V)=e^{-\{\Gamma V\}}{\mathcal R}V+
\frac{1-e^{-\{\Gamma V\}}}{\{\Gamma V\}}{\mathcal N}V-V.
~~\Box
$$
\\The control term can be expanded in power series as
\bq
f(V)=\sum_{n=1}^{\infty}\frac{(-1)^n}{(n+1)!}\{\Gamma V\}^n
(n{\mathcal R}+1)V.
\label{expansion_f}
\eq
\\
We notice that
if $V$ is of order $\epsilon$, $f(V)$ is of order $\epsilon^2$.\\
\indent Proposition 1 tells that the addition of a well chosen 
control term $f$ makes the Hamiltonian canonically
conjugate to $H_0+{\mathcal R} V$. It is also possible to 
show from Proposition 1
that the flow of $H_0+V+f$ is conjugate to the flow of $H_0+{\mathcal R}V$
(see Ref.~\cite{michel}).
\begin{proposition} --
$$
\forall {t} \in \bbR , \qquad e^{{t} \{{H_0} + {V} + {f}\}} =
e^{-\{\Gamma {V}\}} ~ e^{{t} \{{H_0}\}}~ e^{{t} \{\mathcal R {V}\}}
~ e^{\{\Gamma {V}\}}. 
$$
\end{proposition}
The remarkable fact is that
the flow of ${\mathcal R}V$ commutes with the one of $H_0$, since
$\{H_0\}{\mathcal R} = 0$. This allows the splitting of the flow of 
$H_0 +{\mathcal R} V$ into a product. \\ \indent The notion of 
{\em non-resonant} Hamiltonian is defined by the following statement:
\\ 
{\bf Definition} -- ~~
$H_0$ is {\em non-resonant} if and only if
$
\forall {\bf A}\in{\mathcal B}, {\bm \omega}({\bf A})\cdot{\bf k}=0 
$
implies ${\bf k}={\bf 0}$.\\
\indent If $H_0$ is non-resonant then with the addition of a
control term $f$, the Hamiltonian $H_0+V+f$ is 
canonically conjugate to the integrable Hamiltonian $H_0+{\mathcal R} V$
since ${\mathcal R} V$ is only a function of the 
actions [see Eq.~(\ref{eqn:RV})].
\\ \indent If $H_0$ is resonant and ${\mathcal R} V=0$, the controlled
Hamiltonian $H=H_0+V+f$ is conjugate to $H_0$.
\\ In the case ${\mathcal R} V=0$, the series (\ref{expansion_f})
which gives the expansion of the control term $f$,
can be written as
\bq
f(V)=\sum_{s=2}^{\infty}f_s,
\label{exp_f_rv_0}
\eq
where $f_s$ is of order $\epsilon^s$ and given by the
recursion formula
\bq
f_s=-\frac{1}{s}\{\Gamma V,f_{s-1}\},
\label{recursion}
\eq
where $f_1=V$.

\noindent {\em Remark :} A different approach of control has been
developed by G. Gallavotti in
Ref.~\cite{gallavotti}. The idea is to find a
control term (named {\it counter term}) depending only on the actions,
i.e.\ to find $N$ such that
$$
H({\bf A},{\bm \varphi})=H_0({\bf A})+V({\bf A},{\bm \varphi})-N({\bf A}),
$$
is integrable. For isochronous systems, that is
$$
H_0({\bf A})={\bm \omega}\cdot {\bf A},
$$
or any function $h({\bm \omega}\cdot {\bf A})$,
it is shown that if the frequency vector satisfies a Diophantine
condition and if the perturbation is sufficiently small and smooth, such
a control term exists. An algorithm to compute it by recursion
is provided by the proof. We notice that the resulting control term $N$
is of the same order as the perturbation, and has the following
expansion
$$
N({\bf A})={\mathcal R} V+\frac{1}{2} {\mathcal R} \{\Gamma V \}
V+O(\varepsilon^3),
$$
where we have seen from Eq.~(\ref{eqn:RV}) that ${\mathcal R}V$ is only
a function of the actions in the non-resonant case.
The assumption that ${\bm \omega}$
is non-resonant is a crucial hypothesis in Gallavotti's
renormalization approach. Otherwise, a counter-term which only depends
on the actions ${\bf A}$ cannot be found.\\ 
Our approach makes possible the
construction of a control term in the resonant case. The controlled 
Hamiltonian is conjugate to $H_0+{\mathcal R} V$ where ${\mathcal R} V$
depends on the angle and action variables in the resonant case.
Therefore the controlled Hamiltonian is not integrable in general. 
The new term ${\mathcal R} V$ which is always a conserved quantity is
functionally independent of $H_0$ since it depends on the angles.
There exists a linear canonical transformation 
 $({\bf A}^{\prime},{\bm \varphi}^{\prime})=(^tT{\bf A},T^{-1}{\bm \varphi})$ 
where $T$ is a
$l\times l$ matrix with integer coefficients and determinant $1$
such that ${\bm \omega}$ is mapped onto a new frequency vector which has its
$r$ last components equal to zero, where $r$ denotes the dimension of
$\{{\bf k}\in {\mathbb Z}^l \mbox{ s.t. }{\bm \omega}\cdot{\bf k}=0\}$. 
In these new
coordinates ${\mathcal R} V$ depends only on $r$ angles. This form of
$H_0+{\mathcal R} V$ is called the {\it resonant normal form}.
The non resonant case occurs when $r=0$. When $r=1$ the normal form of
$H_0+{\mathcal R} V$ depends only on one angle, so it is integrable.\\
In what follows, we will apply the control theory to a resonant Hamiltonian
which models the ${\bf E}\times{\bf B}$ drift in magnetized plasmas.
\section{Control of chaos in a model for ${\bf E}\times{\bf B}$ 
drift in magnetized plasmas}
\label{sec:3}
\subsection{The model}
\label{sec:3a}
In the guiding centre approximation, the equations of motion
of a charged particle in presence of a strong toroidal magnetic field
and of a time dependent electric field are \cite{Northrop}
\bq
{\dot{\bf x}}= \frac{d}{dt}{x \choose y}=\frac{c}{B^2}{\bf E}({\bf
x},t)\times {\bf B}= \frac{c}{B}{-\partial_y V (x,y,t)\choose
\partial_x V (x,y,t)},
\label{guidcent}
\eq
where $V$ is the electric potential, ${\bf E}=-{\bf\nabla} V$,
and ${\bf B}=B {\bf e}_z$. 
The spatial coordinates $x$ and $y$ where $(x,y)\in \mathbb{R}^2$ play the role of the canonically
conjugate variables and the electric potential $V(x,y,t)$
is the Hamiltonian of the problem.
To define a model we choose
\bq
V ({\bf x},t)=\sum_{{\bf k}\in{\mathbb Z^2}}{V_{\bf k}}\sin \left [ \frac{2\pi}{L}{\bf k}\cdot {\bf x}
+\varphi_{\bf k}-\omega ({\bf k})t\right ],
\eq
where $\varphi_{\bf k}$ are random phases (uniformly distributed)
and $V_{\bf k}$ decrease as
a given function of $\vert {\bf k}\vert$, in agreement with experimental data
\cite{anormal_exp}.
In principle one should use for $\omega ({\bf k})$ the dispersion
relation for electrostatic drift waves (which are thought to be
responsible for the observed turbulence) with a frequency broadening
for each ${\bf k}$ in order to model the experimentally observed
spectrum $S({\bf k},\omega)$. In order to use a simplified model we use
in this article  $\omega({\bf k})=\omega_0$ constant
as a dispersion relation.
The phases $\varphi _{\bf k}$ are chosen at
random in order to mimic  a turbulent field with the reasonable hope 
that the properties of the realization thus obtained are not significantly 
different from their average. 
In addition we take for $|{V_{\bf k}}|$ a power law in
$|{\bf k}|$ to reproduce the spatial spectral characteristics of the
experimental $S({\bf k})$, see Ref.~\cite{anormal_exp}. Thus
we consider the following explicit form of the electric potential
\bQ
V (x,y,t) &=&\frac{a}{2\pi}\sum_{m,n=1\atop{n^2+m^2\le N^2}}^N
\frac{1}{(n^2+m^2)^{3/2}}\times\nonumber\\
&&\sin \left[\frac{2\pi}{L}(nx + my) +
\varphi_{nm} -\omega_0 t \right] .\nonumber\\
\label{potential}
\eQ
By rescaling space and time, we can always assume that 
$L=1$ and $\omega_0=2\pi$.
In what follows, we choose $N=25$. 
\begin{figure}
\epsfig{file=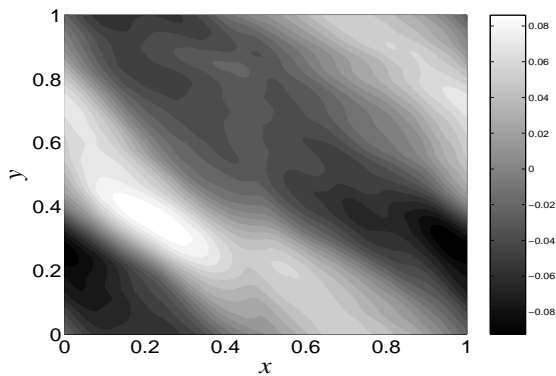,width=7.5cm,height=5.0cm}
\caption{Contour plot of $V(x,y,t)$  given by Eq.(\ref{potential})
for  $t=0$, $a=1$ and $N=25$.}
\label{figurepot}
\end{figure}
Figure~\ref{figurepot} shows a visualization of the potential 
for $t=0$ and $a=1$. 
Since the model is fluctuating in time, the eddies of 
Fig.\ \ref{figurepot} are rapidly modified in 
time and where a vortex was initially present, 
an open line appears, and so on.\\
Two particular properties of the model, anisotropy and propagation
have been observed : each image of the potential field shows an elongated
structure of the eddies and superposing images obtained
at different times a slight propagation in the $y=x$ direction is found.
However, this propagation can easily be proved not to disturb
the diffusive motion of the guiding centers. The property of propagation
can be easily understood analytically. In fact, restricting ourselves
to the most simplified case of an electric potential given only by a dominant
mode ($n=m=1$) it is immediately evident that at any given time the maxima
and minima of the sine are located on the lines $y=-x+constant$.
As the amplitudes are decreasing functions of $n$ and $m$, 
this structure is essentially preserved also in the case of many waves.
The property of anisotropy is an effect of the random phases 
in producing eddies that are irregular in space.\\
\indent We notice that there are
two typical time scales in the equations of motion: the drift characteristic 
time $\tau_d$, inversely proportional to the parameter $a$,
and the period of oscillation $\tau_{\omega}$ of all the waves 
that enter  the potential. The competition between these two time scales
determines what kind of diffusive behaviour is observed~\cite{marc88}.
In what follows we  consider the case of 
weak or intermediate chaotic dynamics 
(coexistence of ordered and chaotic trajectories) which corresponds to the
quasi-linear diffusion regime~(see Sec.~\ref{diffusion of test particles}).
Whereas in the case of fully developed chaos that corresponds 
to the so-called Bohm diffusion regime one has to introduce a slightly
more complicated approach (see remark at the end of Sec.~\ref{sec:3c}).
\subsection{Computation of the control term}
\label{sec:3b}
We extend the phase space $(x,y)$ into
$(x,y,E,\tau)$ where the new dynamical variable $\tau$ evolves as 
$\tau(t)=t+\tau(0)$ and $E$ is its canonical conjugate. The 
autonomous Hamiltonian of the model is
\bq
H(x,y,E,\tau) = E + V(x,y,\tau)\label{hamilton}.
\eq
The equations of motion are
\bq
{\dot x}=-\frac{\partial H}{\partial y} =
-\frac{\partial V}{\partial y},\ \ \ \ \ {\dot y}=\frac{\partial
H}{\partial x} = \frac{\partial V}{\partial x},\ \ \ \ \ {\dot \tau}= 1,
\label{Hequations}
\eq
and $E$ is given by taking $H$ constant along the trajectories.
We absorb
the constant $c/B$ of Eq.~(\ref{guidcent}) in the amplitude $a$ of
Eq.~(\ref{potential}), so that
we can assume that $a$ is small  when $B$ is large.
Thus, for small values of $a$, Hamiltonian 
(\ref{hamilton}) is in the form $H=H_0+\epsilon V$, 
that is an integrable 
Hamiltonian $H_0$ (with action-angle variables)
plus a small perturbation $\epsilon V$.
In our case $H_0=E$, i.e. independent of $x,y,\tau$, so that ${\bf
A}=(E,x)$ and ${\bm\varphi}=(\tau,y)$ are action-angle coordinates for
$H_0$ ($y$ can be considered as an angle but it is frozen by the flow of
$H_0$). We could have exchanged the role of $x$ and $y$. We have
$$
\bm\omega({\bf A})=\left(\frac{\partial H_0}{\partial E},\frac{\partial
H_0}{\partial x}\right)=(1,0),
$$
that is $H_0$ is resonant [i.e. $\bm\omega({\bf A})\cdot{\bf k}=0$ does not
imply ${\bf k=0}$ ]. 
In order to construct the operators $\Gamma$, $\mathcal R$ and 
$\mathcal N$ we consider
$$
\frac{\partial}{\partial{\bm \varphi}}=\left(\frac{\partial}{\partial\tau},
\frac{\partial}{\partial y}\right),
$$
so 
$$
\{H_0\}=\bm\omega({\bf A})\cdot\frac{\partial}{\partial\bm\varphi}
 =\frac{\partial}{\partial{\tau}}.
$$
If we consider an element $W(x,y,\tau)$ of the algebra 
$\mathcal A$, periodic in time with period 1, we can write
$$
W(x,y,\tau)=\sum_{k\in\mathbb Z}W_{k}(x,y)e^{2i\pi k\tau},
$$
and the action of $\Gamma$, $\mathcal R$ and $\mathcal N$ 
operators on $W$ is given by 
\bq
\label{eqn:gamma}
\Gamma W=\sum_{k\neq 0}\frac{W_{k}(x,y)}{2ik\pi }e^{2i\pi k\tau},
\eq
$$
{\mathcal R} W=W_0(x,y),
$$
$$
{\mathcal N} W=\sum_{k\neq 0}W_{k}(x,y)e^{2i\pi k\tau}.
$$
\begin{figure}
\epsfig{file=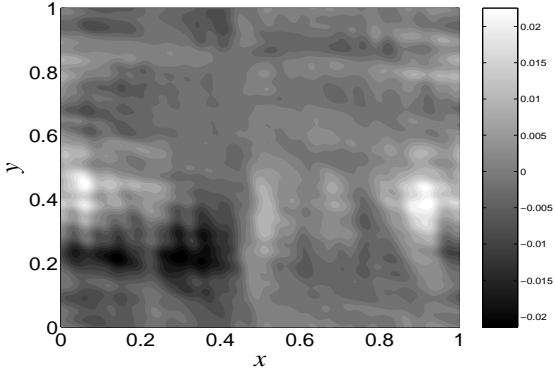,width=7.5cm,height=5.0cm}
\caption{Contour plot of $f_2$ given by Eq.~(\ref{f_2}) for $a=1$ and $N=25$.}
\label{figuref2}
\end{figure}
If we apply the operator $\Gamma$ to $V$ 
given by Eq.~(\ref{potential}), we obtain
\bQ
\Gamma V&=&\frac{a}{(2\pi)^2}
\sum_{n,m=1\atop{n^2+m^2\leq N^2}}
\frac{1}{(n^2+m^2)^{3/2}}\times\nonumber\\
&&\cos\left[2\pi(nx+my)+\varphi_{nm}-2\pi \tau\right].
\label{GammaV}
\eQ
Since $V$ is periodic in time with zero mean value, we have  
${\mathcal R}V=0$. 
In this case, as we have seen in the previous section,
Eqs.~(\ref{exp_f_rv_0})-(\ref{recursion})  give the expansion of the 
control term. 
If we add the exact expression of the control term to $H_0+V$, 
the effect on the flow is the confinement of the motion,
i.e. the fluctuations of the trajectories of the particles, around
their initial positions, are uniformly bounded for any time~\cite{michel}.\\
In the present article we show that truncations of the exact control
term $f$ are able to regularize the dynamics and to slow down the diffusion.
We compute the
first terms of the series of the exact control term $f_2$ and $f_3$. 
From Eq.~(\ref{recursion}) we have
$$
f_2(V)=-\frac{1}{2}\{\Gamma V,V\},
$$
and using the expressions of $V$ and $\Gamma V$, we have
\begin{eqnarray*}
&&f_2({\bf x},\tau)=-\frac{a^2}{2(2\pi)^3}\sum_{n_1,m_1\atop{n_2,m_2}}
\frac{1}{(n_1^2+m_1^2)^{3/2}(n_2^2+m_2^2)^{3/2}}\times\\ 
&&\{\cos(2\pi(n_1x+m_1y)+\varphi_{n_1m_1}-2\pi\tau),\\ 
&&~~~~~~~~~~~\sin(2\pi(n_2x+m_2y)+\varphi_{n_2m_2}-2\pi\tau)\},
\end{eqnarray*}
where $\{\cdot,\cdot\}$ is the Poisson bracket for $x,y$ coordinates, i.e.
for two generic functions $f$ and $g$ depending on $x,y,\tau$,
we have
$$
\{f,g\}=\frac{\partial f}{\partial x}\frac{\partial g}{\partial y}-
\frac{\partial f}{\partial y}\frac{\partial g}{\partial x}.
$$
From Eqs.~(\ref{potential}) and (\ref{GammaV})   we obtain
\bQ
&&f_{2}(x,y,\tau)=
\frac{a^2}{8\pi}\sum_{n_1,m_1\atop{n_2,m_2}} \frac{n_1 m_2 - n_2 m_1}{ 
 (n_1^2+m_1^2)^{3/2} (n_2^2+m_2^2)^{3/2}}  \nonumber\\
&&\times\sin \bigl[ 2\pi \bigl[ (n_1-n_2) x + (m_1-m_2) y\bigr] +
\varphi_{n_1 m_1} - \varphi_{n_2 m_2} \bigr] .  \nonumber
\label{f_2}\\
\eQ
We notice that for the particular model
(\ref{potential}) and for the particular choice of operator $\Gamma$ given by 
Eq.~(\ref{eqn:gamma}), the partial control term
$f_2$ is independent of time. Figure~\ref{figuref2} depicts a contour plot of it.
\\ \indent The computation of $f_3$ is given by
$$
f_3(V)=-\frac{1}{3}\{\Gamma V,f_2\},
$$
and substituting the expressions (\ref{GammaV}) and (\ref{f_2}) for $\Gamma V$
and $f_2$, one obtains
\bQ
&&f_3(x,y,\tau)=-\frac{a^3}{24\pi}\times\nonumber\\
&&\sum_{n_1,m_1,n_2,m_2\atop{n_3,m_3}}
\frac{(n_1m_2-m_1n_2)\left[(n_1-n_2)m_3-(m_1-m_2)n_3\right]}
{(n_1^2+m_1^2)^{3/2}(n_2^2+m_2^2)^{3/2}(n_3^2+m_3^2)^{3/2}}\nonumber\\ 
&&\times\sin[2\pi(n_1-n_2+n_3)x+2\pi(m_1-m_2+m_3)y+\nonumber\\
&&~~~~~~~~~~\varphi_{n_1m_1}-
\varphi_{n_2m_2}+\varphi_{n_3m_3}-2\pi\tau)] .
\label{f_3}
\eQ
\begin{figure}
\epsfig{file=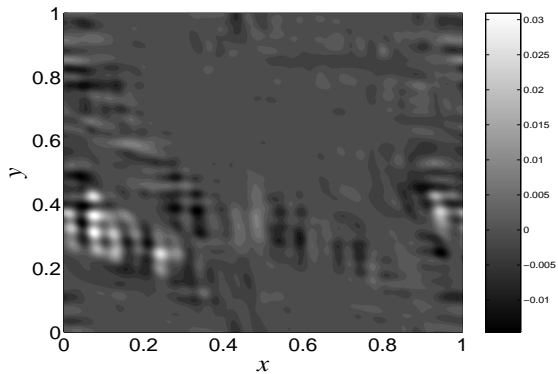,width=7.5cm,height=5.0cm}
\caption{Contour plot of $f_3$ given by Eq.~(\ref{f_3}) for $t=0$, $a=1$ and $N=25$.}
\label{figuref3}
\end{figure}
The computation of the other terms of the series (\ref{exp_f_rv_0})
can be done recursively by using Eq.(\ref{recursion})
(see also \cite{michel} and the Appendix).

\subsection{Properties of the control term}
\label{sec:3c}
In this section, we first state that for $a$ sufficiently small, 
the exact control term exists and is regular. The proofs of these propositions are given in the Appendix. Then we give estimates of the partial 
control terms in order to compare the relative sizes of the different 
terms with respect to the perturbation.\\
Concerning the existence of the control term, we have the following proposition
\begin{proposition}
\label{proposition3} -- If the amplitude $a$ of the potential 
is sufficiently small, there exists a control term $f$ given 
by the series (\ref{exp_f_rv_0}) such that $E+V+f$ is canonically conjugate 
to $E$, where $V$ is given by Eq.~(\ref{potential}).
\end{proposition}
The proof is given in the Appendix. For $N=25$, it is shown that the control term exists for $a\lesssim 7\times 10^{-3}$. As usual, such estimates are very conservative with respect to realistic values of $a$. In the numerical study, we consider values of $a$ of order 1.

Concerning the regularity of the control term, we notice that each term $f_s$ in the series~(\ref{exp_f_rv_0}) is a trigonometric polynomial with an increasing degree with $s$. The resulting control term is not smooth but its Fourier coefficients exhibits the same power law mode dependence as $V$:
\begin{proposition}
\label{proposition4} -- All the Fourier coefficients $f_{nmk}^{(s)} $ of the functions $f_s$ of 
the series~(\ref{exp_f_rv_0}) satisfy:
$$
\vert f_{nmk}^{(s)}\vert  \leq  \frac{a^s C^s}{(n^2+m^2)^{3/2}},
$$ 
for $(n,m)\not= (0,0)$. Consequently, for $a$ sufficiently small, the Fourier coefficients of the control term $f$ given by Eq.~(\ref{exp_f_rv_0}) satisfy:
$$
\vert f_{nmk}\vert  \leq  \frac{C_\infty}{(n^2+m^2)^{3/2}},
$$
for $(n,m)\not= (0,0)$ and for some constant $C_\infty >0$.
\end{proposition}

In order
to measure the relative magnitude between Hamiltonian
(\ref{potential}) and $f_2$ or $f_3$, we have numerically computed their mean
squared values:
\begin{eqnarray*}
&&\sqrt{\frac{\langle f_{2}^{2}\rangle} 
 {\langle V^2\rangle}} \approx 0.13 a,\\
&& \sqrt{\frac{\langle f_{3}^{2}\rangle} 
 {\langle V^2\rangle}} \approx 0.07 a^2,
\end{eqnarray*}
where $\langle f \rangle=\int_0^{1} dt\int_0^1dx\int_0^1dy f(x,y,t)$.\\

Another measure of the relative sizes of the control terms is 
by the electric energy density associated with each electric field 
$V$, $f_2$ and $f_3$. From the potential we get the electric field and hence 
the motion of the particles. We define an average energy 
density ${\mathcal E}$ as
$$
{\mathcal E}=\frac{1}{8\pi}\langle~ |~{\bf E}~|^2~\rangle
$$
where ${\bf E}(x,y,t)=-{\bf\nabla}V$. In terms of the particles, it corresponds to the mean value of the kinetic energy $\langle \dot{x}^2+\dot{y}^2\rangle$ (up to a multiplicative constant). For $V(x,y,t)$ given by 
Eq.~(\ref{potential}),
\begin{equation}
{\mathcal E}=\frac{a^2}{8\pi}\sum_{n,m=1\atop{n^2+m^2\le N^2}}^N
\frac{1}{(n^2+m^2)^2}.
\label{enepotenziale}
\end{equation}
We define the contribution of $f_2$ and $f_3$ to the energy density by
\begin{eqnarray}
&& e_2=\frac{1}{8\pi} \langle \vert {\bf \nabla}f_2 \vert^2\rangle, \label{e2}\\
&& e_3=\frac{1}{8\pi} \langle \vert {\bf \nabla}f_3 \vert^2\rangle.
\end{eqnarray}
For $N=25$, these contributions satisfy:
$$
\frac{e_2}{{\mathcal E}}\approx 0.1\times a^2, \qquad \frac{e_3}{{\mathcal E}}\approx 0.3\times a^4.
$$
It means that the control terms $f_2$ and $f_3$ can be considered as small
perturbative terms with respect to $V$ when $a<1$. We notice that even 
if $f_3$ has a smaller amplitude than $f_2$, its associated average energy density 
is larger for $a$ of order 1 (more precisely for $a\geq 0.58$). \\

{\em Remark on the number of modes in $V$}: In Sec.\ref{sec:3d}, all the computations have been performed for a fixed number of modes $N$ ($N=25$) in the potential $V$ given by Eq.~(\ref{potential}). The question we address in this remark is how the results are modified as we increase $N$. First we notice that the potential and its electric energy density are bounded with $N$ since
\begin{eqnarray*}
&& \vert V(x,y,t)\vert \leq a \sum_{n,m=1}^\infty \frac{1}{(n^2+m^2)^{3/2}} < \infty ,\\
&& {\mathcal E}\leq \frac{a^2}{8\pi}\sum_{n,m=1}^\infty
\frac{1}{(n^2+m^2)^2}< \infty .
\end{eqnarray*}
Concerning the partial control term $f_2$, we see that it is in general unbounded with $N$. From its explicit form, one can see that it grows like 
$N\log N$ (see the Appendix).
Less is known on the control term since it is given by a series whose terms are defined by recursion. 
However, from the proof of Proposition~\ref{proposition3} in the Appendix, we see that the value $a$ of existence of the control term decreases like 
$1/(2N\log N)$. This divergence of the control term comes from the fact that the Fourier coefficients of the potential $V$ are weakly decreasing with the amplitude of the wavenumber.\\
Therefore, the exact control term might not exist if we increase $N$ 
keeping $a$ constant. 
However we will see in Sec.~\ref{sec:5b} that for practical purposes 
the Fourier series of 
the control term can be truncated to its first terms 
(the Fourier modes with highest amplitudes). 
Furthermore in the example we considere as 
well as for any realistic situation the value of $N$ is bounded by the 
resolution of the potential. In the case of electrostatic turbulence in plasmas
$k\rho_i\sim 1$  determines an upper bound for $k$, where $k$ is the 
transverse wave vector related to the indices $n,m$
and $\rho_i$ the ion Larmor radius,. 
The physics corresponds to the averaging effect introduced 
by the Larmor rotation.\\

{\em Remark on the control in the Bohm regime (parameter $a$ larger than 1):}
Let us define $V_0\equiv V(t=0)$ and $\delta V=V-V_0$. Since the Bohm regime is
defined as a regime of relatively slow evolution of the potential 
(with characteristic time $\tau_{\omega}$)
compared to the motion of the particles 
(with characteristic time $\tau_{d}$),
 $\tau_d<\tau_{\omega}$, one can introduce
as small parameter $\epsilon\sim\tau_d/\tau_{\omega}$ and the 
integrable Hamiltonian $\bar H_0=H_0+V_0$, so that 
$H=\bar H_0+\epsilon \tilde V$ with $\tilde V=(V-V_0)/\epsilon$. This approach
is only valid for a finite time of order $\tau_{\omega}$ after which one
must redefine $V_0$, $\bar H_0$ and $\tilde V$. 
\section{Numerical investigation of the control term}
\label{sec:3d}
With the aid of numerical 
simulations (see Ref.~\cite{marc88}
for more details on the numerics), we check the effectiveness of
the control theory developed in Sec.~\ref{sec:3}  
by comparing the dynamics of 
particles  obtained from Hamiltonian (\ref{potential}) and
from the same Hamiltonian with the control term $f_2$ 
given by Eq.~(\ref{f_2}), 
and with a more refined control term $f_2+f_3$ where $f_3$ is given by 
Eq.~(\ref{f_3}). We use three types of indicators of the dynamics: diffusion 
coefficient, Lyapunov indicators, and probabilty distribution function (PDF)
of step sizes.

\subsection{{\bf Diffusion of test particles}}
\label{diffusion of test particles}
The effect of the control terms can first be seen from a few randomly chosen 
trajectories. We have plotted Poincar\'e sections (which are stroboscopic 
plots with period $1$ of the trajectories of $V$). 
On Figures \ref{figure1} and \ref{figure2} is plotted 
the Poincar\'e surfaces of section of two
trajectories issued from generic initial conditions computed without
and with the control term $f_2$ respectively. Similar pictures are obtained 
for many other randomly chosen initial conditions.
The stabilizing effect of the control term (\ref{f_2}) is illustrated
by such trajectories. The motion remains diffusive
but the extension of the phase space explored by the trajectory is reduced.
\begin{figure}
\caption{Poincar\'e surface of section of a trajectory obtained for
Hamiltonian (\ref{potential}) using a generic initial condition
assuming $a=0.8$.}
\label{figure1}
\end{figure}
\begin{figure}
\caption{Poincar\'e surface of section of a trajectory obtained 
using a generic initial condition as in Fig.1 and adding the control term 
(\ref{f_2}) to Hamiltonian (\ref{potential}) with $a=0.8$.}
\label{figure2}
\end{figure}
\\ \indent In order to study the diffusion properties of the system, we have
considered a set of $\mathcal M$ particles (of order $1000$) 
uniformly distributed at
random in the domain $0\leq x,y\leq 1$ at $t=0$. We have computed the
mean square displacement $\langle r^2 (t)\rangle$ as a function of
time
\bq
\langle r^2 (t)\rangle = \frac{1}{\mathcal M} \sum_{i=1}^{\mathcal M}
{|{\bf x}_i(t) - {\bf x}_i(0)|}^2,
\eq
where ${\bf x}_i(t)=(x_i(t),y_i(t))$ is the position of the
$i$-th particle at time $t$ obtained by integrating Eq.~(\ref{Hequations}) 
with initial condition ${\bf x}_i(0)$.
On Figure~\ref{figure3} is presented $\langle r^2 (t) \rangle$ for three different values of
$a$: $a=0.7$, $a=0.8$ and $a=0.9$.
\begin{figure}
\epsfig{file=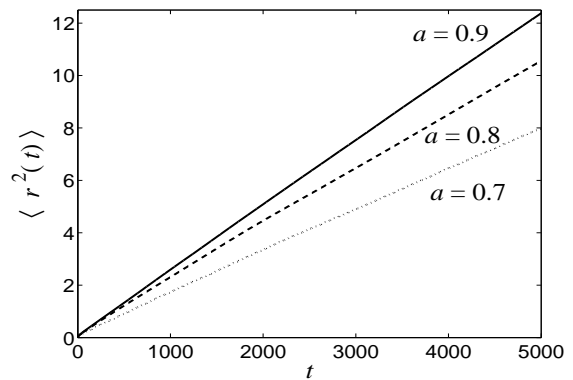,width=7.5cm,height=5.0cm}
\caption{Mean square displacement $\langle r^2 (t)\rangle $ versus
time $t$ obtained for  Hamiltonian
(\ref{potential}) with three different values of $a=0.7,0.8,0.9$.}
\label{figure3}
\end{figure}
For the range of parameters we consider the behavior 
of $\langle r^2 (t)\rangle $ is always found to be 
linear in time for $t$ large enough. The
corresponding diffusion coefficient is defined as
$$
D= \lim_{t \rightarrow\infty} \frac{\langle r^2(t)\rangle}{t}.
$$
The values of $D$ as a function of $a$ with and without 
control term are presented on Figure~\ref{figure4}. 
A significant decrease of the diffusion 
coefficient when the control term $f_2$ is added can be readily be observed.
As expected, the action of the control term gets weaker as $a$ is
increased towards the strongly chaotic region.
\begin{figure}
\epsfig{file=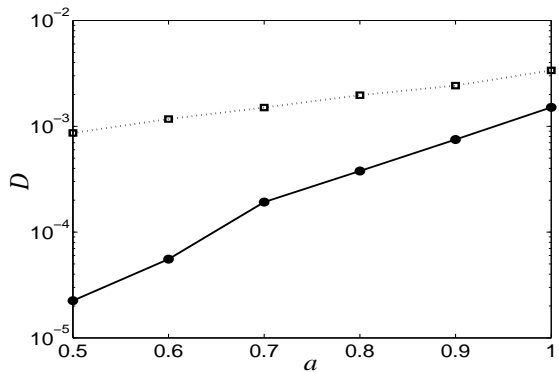,width=7.5cm,height=5.0cm}
\caption{Diffusion coefficient $D$ versus $a$ 
obtained for Hamiltonian (\ref{potential}) (open squares) and Hamiltonian
(\ref{potential}) plus control term (\ref{f_2}) (full
circles).}
\label{figure4}
\end{figure}
\subsection{{\bf Lyapunov indicator method}}

In order to get insight into the action of the control term to the
dynamics, we 
apply the Lyapunov Indicator method.
This method
provides local information in phase space. It has been introduced to
detect ordered
and chaotic trajectories in the set of initial conditions. It
associates a finite-time Lyapunov exponent $\nu$
with an initial condition ${\bf x}_0$. By looking at the map ${\bf x}_0
\mapsto \nu({\bf x}_0)$, one distinguishes the set of initial
conditions leading to regular motion associated with a small 
finite-time Lyapunov exponent. The pictures of
this map show the phase space structures where the motion is trapped
and does not diffuse throughout phase space, e.g., they highlight islands
of stability located around elliptic periodic orbits. \\

\begin{figure}
\epsfig{file=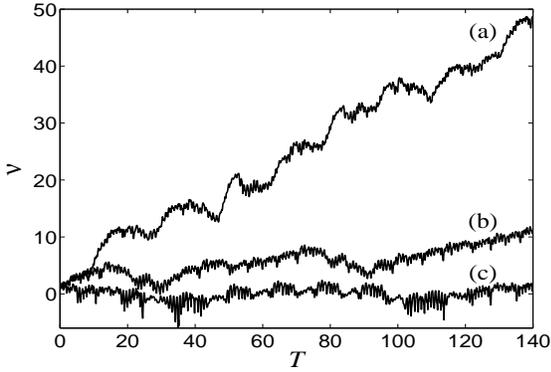,width=7.5cm,height=5.0cm}
\caption{Values of the Lyapunov indicator $\nu({\bf x_0},T)$ as a function of time
$T$ for three trajectories obtained for Hamiltonian~(\ref{potential})
with $a=0.4$ for the following initial conditions: one strongly chaotic
$(a)$ ${\bf x}_0=(0.865, 0.39)$, one weakly chaotic $(b)$ ${\bf
x}_0=(0.8766, 0.39)$, and one quasi-periodic $(c)$ ${\bf x}_0=(0.895,
0.39)$.}
\label{fig:fli1}
\end{figure}

\begin{figure}
\epsfig{file=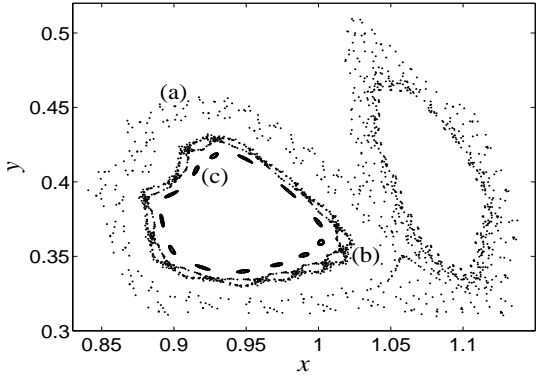,width=7.5cm,height=5.0cm}
\caption{Poincar\'e sections of the three trajectories of
Fig.~\ref{fig:fli1}.}
\label{fig:fli2}
\end{figure}

Consider an autonomous flow $\dot{{\bf x}}=f({\bf x})$. The Lyapunov
indicator method is based on the analysis of the tangent flow 
\begin{equation}
\frac{d {\bf y}}{dt}= Df({\bf x}) {\bf y},
\label{tangflow}
\end{equation}
where $Df$ is the matrix of variations of the flow. The Lyapunov indicator is
defined as the value  
$$
\nu({\bf x}_0,T)=\log \Vert {\bf y}(T)\Vert,
$$ 
at some finite-time $T$ starting with some initial condition ${\bf x}_0$
and a generic vector ${\bf y}_0$. This definition is very close to the
one of a finite-time Lyapunov exponent. The plot of $\nu$ versus ${\bf
x}_0$ gives a map of the dynamics by highlighting regions of stability
and regions of chaotic dynamics.\\
For a chaotic trajectory, the value of the Lyapunov indicator increases
linearly with time, whereas for a regular trajectory (periodic or
quasi-periodic), it increases like $\log t$ (see rigorous results for
nearly perturbed Hamiltonian systems in Ref.~\cite{guzz02}). So in regular regions, this Lyapunov indicator is expected to be much lower than in chaotic regions.\\
Here the Hamiltonian flow is not autonomous. However, by considering
that $t$ is a new coordinate of the motion (and $E$ is its conjugate
momentum), we obtain an autonomous flow with two degrees of freedom. We
notice that the equations of motion for Hamiltonian~(\ref{potential})
can be written as 
\begin{equation}
\dot{{\bf x}}= {\rm Re}\left[ {\bf F}({\bf x}) e^{-2i\pi t}\right],
\label{tgflow2}
\end{equation}
where 
$$
 {\bf F}({\bf x})={\bf \nabla}^{\perp}V({\bf x},0)+i{\bf \nabla}^{\perp}V({\bf x},1/4),
$$
where ${\bf \nabla}^{\perp}=(-\partial/\partial y ~, ~ \partial/\partial x)$
and $V({\bf x},t)$ is given by Eq.(\ref{potential}).
The non autonomous flow (\ref{tgflow2}) can be mapped into an autonomous
flow by considering a third equation $\dot\tau=1$. The computation of 
the tangent flow (of dimension $3$) follows from the matrix of variations
of the autonomous flow. We have chosen the third component of the vector
${\bf y}$ following the evolution of the tangent flow (\ref{tangflow}) equals to one,
which can be done without loss of generality since Eq.~(\ref{tangflow})
is linear in ${\bf y}$. Therefore, it reduces to
the evolution of a two dimensional vector ${\bf y}$ which is given by
$$
\dot{{\bf y}} = {\rm Re} \left[ G({\bf x})e^{-2i\pi t}\right] {\bf y}+
2\pi {\rm Im}\left[ {\bf F}({\bf x}) e^{-2i\pi t}\right],
$$
where $G$ is the two-dimensional matrix
$G({\bf x})=D {\bf F}$ (matrix of the variations of the vector 
field ${\bf F}$). 
Figure~\ref{fig:fli1} shows the value of $\nu({\bf x}_0,T)$ as a
function of $T$ for $T\in [0, 140]$ for three initial conditions ${\bf
x}_0$~: one strongly chaotic ${\bf x}_0=(0.865, 0.39)$, one weakly
chaotic ${\bf x}_0=(0.8766, 0.39)$, and one quasi-periodic ${\bf
x}_0=(0.895, 0.39)$. The plot of the Poincar\'e sections of these three
trajectories up to $T=1000$ are shown on Fig.~\ref{fig:fli2}. These
figures show two chaotic trajectories for which there is an overall
linear increase of the Lyapunov indicator, and one quasi-periodic motion
(trapped around an elliptic periodic orbit). We notice that not only the
method is able to discriminate early between regular and chaotic motions
but it is also able to detect weakly versus strongly chaotic
trajectories for rather small values of $T$.\\ 
With the control term $f_2$ given by Eq.~(\ref{f_2}), the
equations of motion can be rewritten as
$$
\dot{{\bf x}}= {\rm Re}\left[ {\bf F} e^{-2i\pi t}\right]+
{\bf f}_2({\bf x}),
$$
where 
$
 {\bf f}_2({\bf x})= {\bf \nabla}^{\perp} f_2.
$
The equation of evolution of ${\bf y}$ becomes
$$
\dot{{\bf y}} =  {\rm Re} \left[ G({\bf x})e^{-2i\pi t}\right] {\bf y}+
2\pi {\rm Im}\left[ {\bf F}({\bf x}) e^{-2i\pi t}\right]+ Df_2({\bf x}){\bf
y}.
$$
With the control term $f_2({\bf x})+f_3({\bf x},t)$ where $f_3$ is given by Eq.~(\ref{f_3}), the equations of motion can be written as
$$
\dot{{\bf x}}= {\rm Re}\left[ \left({\bf F}+{\bf f}_3\right) e^{-2i\pi t}\right]+{\bf f}_2({\bf x}),
$$ 
where 
$
{\bf f}_3({\bf x})={\bf \nabla}^{\perp} f_3({\bf x},0)+i{\bf \nabla}^{\perp} f_3({\bf x},1/4).
$
The equation of evolution of ${\bf y}$ becomes
\begin{eqnarray*}
\dot{{\bf y}} = && {\rm Re} \left[ \left( G+g_3\right)e^{-2i\pi t}\right] {\bf y}\\ &+&
2\pi {\rm Im}\left[ \left({\bf F}+{\bf f}_3\right) e^{-2i\pi t}\right]+ Df_2{\bf
y},
\end{eqnarray*}
where 
$
g_3({\bf x})=D {\bf f}_3.
$

On a grid of 10000 initial conditions ${\bf x}_0\in [0,1]^2$, we compute
$\nu({\bf x}_0,T)$ for $T=200$. Figure~\ref{fig:fli3} represents the
Lyapunov indicator map for $a=0.4$ without control term.
Figure~\ref{fig:fli4} shows for the same values of parameters, the
Lyapunov indicator map with the control term $f_2$. 
\begin{figure}
\caption{Lyapunov indicator map $\nu({\bf x}_0,T=200)$ for 
Hamiltonian~(\ref{potential}) for $a=0.4$.}
\label{fig:fli3}
\end{figure}
\begin{figure}
\caption{Lyapunov indicator map $\nu({\bf x}_0,T=200)$ for 
Hamiltonian~(\ref{potential}) for $a=0.4$ with the control term $f_2$ given
by Eq.~(\ref{f_2}).}
\label{fig:fli4}
\end{figure}
The general effect of the control term $f_2$ is a decrease of the
magnitude of the Lyapunov indicators. However, the stabilization effect
of the control term $f_2$ is not uniform. There are regions where the
(partial) control term $f_2$ fails to stabilize the trajectories, e.g.,
in the region near ${\bf x}_0=(0.1, 0.4)$. Figure~\ref{fig:fli5} plots
the Poincar\'e section of the trajectories starting at ${\bf
x}_0=(0.085, 0.385)$ for Hamiltonian~(\ref{potential}) with and without
the control term $f_2$ given by Eq.~(\ref{f_2}). We notice that this trajectory 
is more chaotic and more diffusive with the control term than without.

\begin{figure}
\epsfig{file=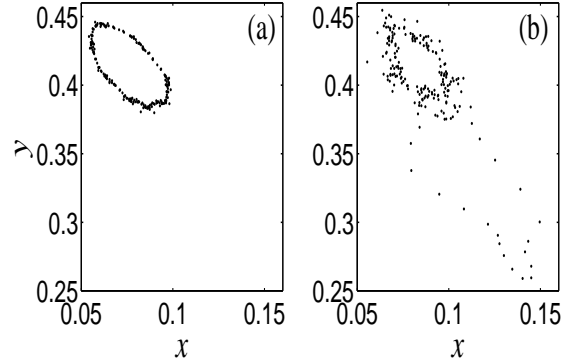,width=7.5cm,height=5.0cm}
\caption{Poincar\'e sections of the trajectories of $(a)$
Hamiltonian~(\ref{potential}) and $(b)$ Hamiltonian~(\ref{potential})
with control term $f_2$ given by Eq.~(\ref{f_2}), with initial conditions ${\bf x}_0=(0.085,
0.385)$.}
\label{fig:fli5}
\end{figure}
\begin{figure}
\epsfig{file=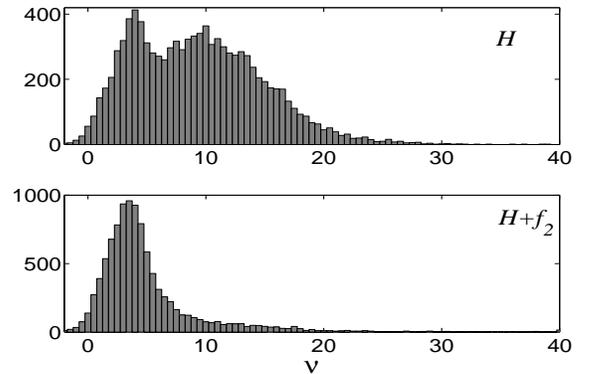,width=7.5cm,height=5.0cm}
\caption{Histograms of the Lyapunov indicators at $T=200$ for $a=0.4$
with and without control term computed in Figs.~\ref{fig:fli3}
and~\ref{fig:fli4}.}
\label{fig:fli6}
\end{figure}
\begin{figure}
\epsfig{file=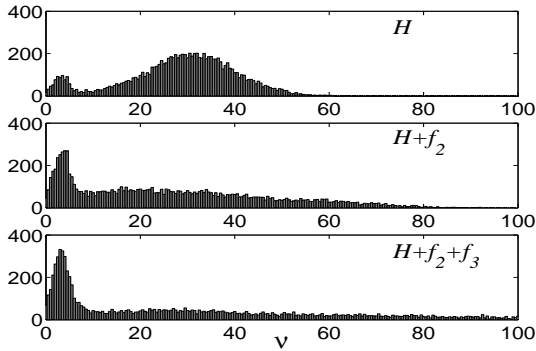,width=7.5cm,height=5.0cm}
\caption{Histograms of the Lyapunov indicators at $T=200$ for $a=0.6$
without control term (upper panel), with control term $f_2$ (middle panel) 
and with control term $f_2+f_3$ (lower panel).}
\label{fig:flif3}
\end{figure}

In order to see the global stabilization effect of the partial control
term $f_2$, we notice from the values plotted in Fig.~\ref{fig:fli3} 
that for $a=0.4$ about 25\% of the trajectories
have a Lyapunov indicator less than 5 at $T=200$ without the control
term $f_2$, compared with 70\% with the control term
(from Fig.~\ref{fig:fli4}).
Figure~\ref{fig:fli6} represents the histograms of the Lyapunov
indicator at $T=200$ for $a=0.4$ with and without control term. The
first peak in the upper and lower panels corresponds to the regular
component of the phase space. We clearly see that the second peak
corresponding to the chaotic component is drastically reduced with the
addition of the partial control term.
The effect of a more refined control term is observed in Fig.~\ref{fig:flif3} for $a=0.6$ when the control term $f_2$ starts to fail to reduce significantly the chaotic part of phase space. 
We see that the proportion of the regular trajectories has increased with the 
addition of $f_3$. More quantitatively, 8\% of the trajectories of the 
Hamiltonian without control have a Lyapunov indicator smaller than 7 
at $T=200$. With the addition of $f_2$, this proportion is increased to 25\% 
whereas it is around 30\% with the addition of $f_3$.  
\subsection{Horizontal step sizes}
\begin{figure}
\epsfig{file=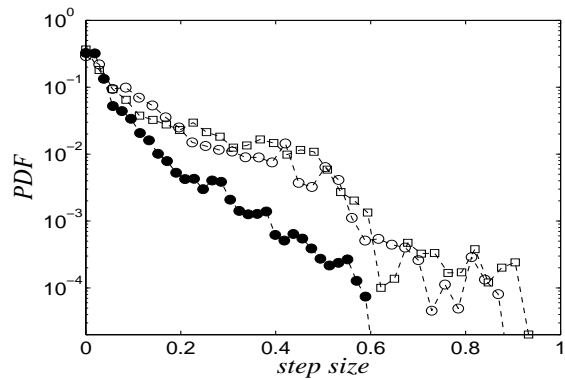,width=7.5cm,height=5.0cm}
\caption{PDF of the magnitude of the horizontal step size 
for Hamiltonian (\ref{potential}) with $a=0.7$
without the control term (open squares), 
with the control term $f_2$ (full circles),
 and with a truncated control term with twelve modes (open circles).}
\label{figure6}
\end{figure}

\begin{figure} 
\epsfig{file=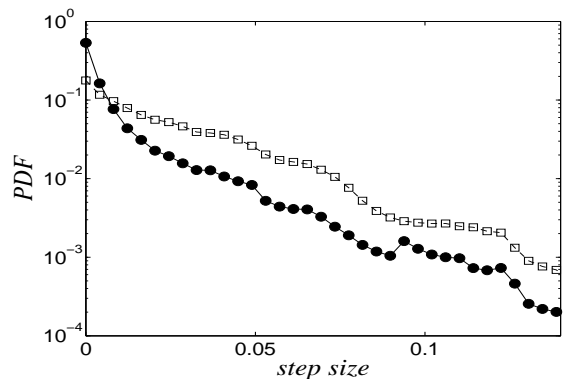,width=7.5cm,height=5.0cm}
\caption{PDF of the horizontal step sizes of the trajectories with small 
Lyapunov indicator (smaller than $7$, full squares) and with large Lyapunov 
indicator (larger than $7$, open squares) 
for Hamiltonian (\ref{potential}) with $a=0.4$.}
\label{PDF_FLI}
\end{figure}

In order to investigate the effect of the control term on the 
transport properties and its relationship with single trajectories, we
have computed the Probability Distribution Function (PDF) of the step sizes.
Let us define the {\em horizontal step size} (resp.~vertical step size) 
as the distance covered by the test particle between two successive sign
reversals of the horizontal (resp.~vertical) component of the drift
velocity. The effect of the control is analyzed in terms
of the PDF of step sizes.
Following test particle trajectories for a large number of initial
conditions, with and without control, leads to the PDFs plotted in
Fig.~\ref{figure6} for Hamiltonian (\ref{potential}) 
without and with control term (\ref{f_2}) for $a=0.7$.
A marked reduction of the PDF is observed at large step
sizes with control relatively to the uncontrolled case.
Conversely, an increase is found for the smaller step sizes. 
The control quenches the large steps (typically larger than $0.5$ for $a=0.7$).
Such a reduction of the probability to achieve a large radial step will modify
the transport efficiency and in particular 
reduce the diffusion coefficient.\\
In order to give support that the first peak of the histogram of 
the Lyapunov indicator (see Fig.~\ref{fig:fli6})
is associated with the small step sizes, we have plotted on the
Fig.~\ref{PDF_FLI} the distribution of
horizontal step sizes of the trajectories with a small Lyapunov indicator
(smaller than $7$), and also the same PDF for trajectories associated 
with large Lyapunov indicator (larger than $7$). This result gives support
to the fact that the control term reduces the diffusion of trajectories
by reducing chaos in the system (by the creation of invariant tori 
\cite{guido1}, see also the global picture in the conclusion).
\section{Robustness of the control}
\label{sec:3e}
In the previous section, we have seen that a truncation of the series 
defining the control term by considering the first or the two first terms
in the perturbation series in $\epsilon$, 
gives a very efficient control on the chaotic dynamics of the system.\\
In this section we show that  it is possible to use an approximate
control or to make a small error while computing the control term and still
get an efficient control of the dynamics.\\
\indent Below, we give numerical evidence for the following statements : 
the reduction of the amplitude shows that one can inject less energy to
achieve a significant control. The truncation of the Fourier series indicate
that one can simplify the control term and still get a significant control.
The change of phases shows that one can introduce some error in the 
phases and still get a significant control.\\

\subsection{Reduction of the amplitude of the control term}
    
\begin{figure}
\epsfig{file=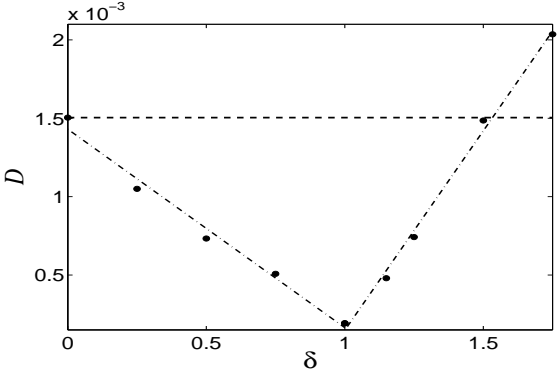,width=7.5cm,height=5.0cm}
\caption{Diffusion coefficient $D$ versus the magnitude of the control term
(\ref{f_2}) for $a = 0.7$. The horizontal dashed line corresponds to the value of $D$
 without control term. The dash-dotted line is a piecewise linear interpolation.}
\label{figure5}
\end{figure}

We check the robustness of the control by increasing or reducing 
the amplitude of the control \cite{guido2}. We replace $f_2$
by $\delta\cdot f_2$ and we vary the parameter $\delta$ away from its
reference value $\delta =1$. Figure~\ref{figure5} shows that both the increase and
the reduction of the magnitude of the control term (which is
proportional to $\delta\cdot a^2$) result in a loss of efficiency in
reducing the diffusion coefficient. The fact that a larger
perturbation term -- with respect to the computed one -- does not work
better, also means that the control is ``smart'' and that it is
not a ``brute force'' effect.\\ The interesting result is that one can 
significantly reduce the amplitude of the control ($\delta <1$) and still 
get a reduction of the chaotic diffusion. We notice that the average energy 
density $e_2(\delta)$ associated with a control term $\delta \cdot f_2$ is 
equal to $e_2(\delta)=\delta^2 e_2$, where $e_2$ is given by Eq.~(\ref{e2}).
Therefore, for $\delta=0.5$ where the energy necessary for the control
is one fourth of the optimal control, the diffusion coefficient is 
significantly smaller than in the uncontrolled case~(nearly factor $3$). 
\subsection{Truncation of the Fourier series of the control term}
\label{sec:5b}
We show that a  reduction of the number of Fourier 
modes of $f_2$ can still significantly reduce chaotic diffusion.
The Fourier expansion of the control term $f_2$ given by Eq.(\ref{f_2}) is
\begin{equation}
f_2=\sum_{n,m}f^{(2)}_{nm}e^{2i\pi(nx+my)}~,
\end{equation}
where $f^{(2)}_{nm}$ is 
\begin{equation}
f^{(2)}_{nm}=\frac{a^2}{8\pi i}\sum_{n_1,m_1}
\frac{(nm_1 - n_1 m)e^{i(\varphi_{n_1m_1}-\varphi_{n_1-n,m_1-m})}}
{(n_1^2+m_1^2)^{3/2}\left[(n_1-n)^2+(m_1-m)^2)\right]^{3/2}}~.
\label{f2_nm}
\end{equation}
The truncation of the Fourier series is made by considering the Fourier
modes with an amplitude greater than or equal to  $\epsilon$, that is 
the sum in Eq.~(\ref{f2_nm}) is restricted to the set of modes $(n,m)$
such that $|f^{(2)}_{nm}|\ge\epsilon$.
\begin{figure}
\epsfig{file=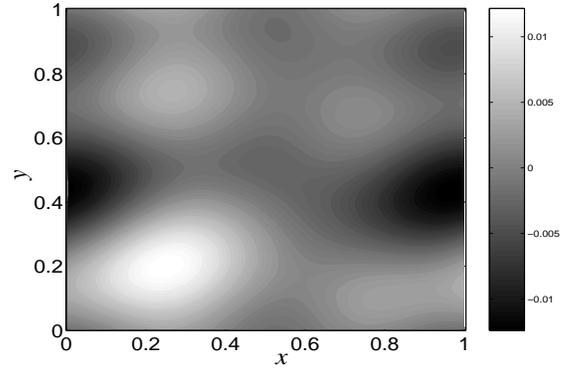,width=7.5cm,height=5.0cm}
\caption{Contour plot of the truncation of $f_2(x,y)$ for $a=0.7$ 
containing the twelve Fourier modes of highest amplitude.}
\label{f2_n6}
\end{figure}
For example, if we consider $\epsilon\simeq 9\cdot10^{-4}$, 
there are twelve modes in the sum, which are the modes with wave vector
$(0,1)$, $(0,2)$, $(1,-1)$, $(1,-2)$, $(1,0)$,
$(2,0)$ and the opposite wave vectors ($f_2$ is real),
compared with the total number of modes of the full $f_2$
which is about $2000$.
Figure \ \ref{f2_n6} shows the contour plot of $f_2$ obtained with 
only these twelve modes.
Figure \ \ref{figureDnmodi} shows the 
diffusion coefficient for the dynamics of the truncated control term
versus the number of Fourier modes kept in the truncation
of $f_2$. The reduction of the diffusion (with 
respect to the uncontrolled case) also holds for a very simplified control
term containing only the few highest Fourier modes
of the full control term. 
\begin{figure}
\epsfig{file=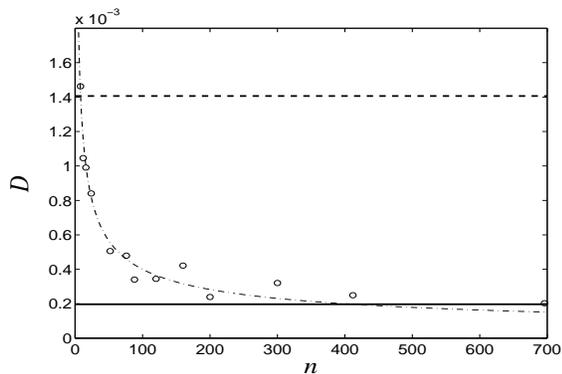,width=7.5cm,height=5.0cm}
\caption{Diffusion coefficient $D$ versus the number of Fourier modes $n$
in the truncation of the control term $f_2$ for $a = 0.7$. 
The dashed line corresponds to the case without control term,
the solid line corresponds to the value of the diffusion with the full
control term $f_2$ and the dash-dotted line corresponds to a power law 
interpolation ($\propto n^{-1/2}$).}
\label{figureDnmodi}
\end{figure}
 If we replace $f_2$ by the truncation with twelve modes
we see that the effect is still a strong reduction of the diffusion
coefficient, a reduction of about $25\%$. 
The energy density 
$e_2^{(12)}$ (see Sec.~\ref{sec:3c}) of this truncated control term with 
respect to the energy density ${\mathcal E}$  is
\begin{equation}
\frac{e_2^{(12)}}{\mathcal E}\simeq0.009\times a^2,
\end{equation}
where $\mathcal E$ is given by Eq.~(\ref{enepotenziale}),
that is less than $1\%$ of the energy associated to 
the electric potential. It is interesting to notice that the energy density
of this truncation with respect to the one of the full control term $f_{2}$
is
\begin{equation}
\frac{e_2^{(12)}}{e_2}\simeq0.09,
\end{equation}
where $e_2$ is given by Eq.~(\ref{e2}),
that is less than $10\%$ of the energy associated with the full control
term $f_2$. Moreover, we see from Fig.~\ref{figure6} where the 
PDF of horizontal step sizes is plotted,
that the effect of this
coarse grained control term $f_2$ reduced to twelve modes 
is also to quench large step sizes.
\\
More generally, these results show that a partial knowledge of the potential,
e.g. on a grid (coarse grained) is sufficient to obtain a significant
control of the dynamics.
\subsection{Change of phases in $f_2$}
We check the robustness of the control with respect to
an error introduced in the phases of the control term 
given by Eq.~(\ref{f_2}), i.e. we change the phases 
$\varphi_{nm}$ by $\tilde\varphi_{nm}$ in Eq.~(\ref{f_2}) by :
\begin{equation}
\tilde\varphi_{nm}=\varphi_{nm}+\gamma\cdot\varphi_{nm}^{\it err}~,
\end{equation}
where $\varphi_{nm}^{\it err}$ are uniformly random distributed 
phases in $[0,2\pi]$ , $\gamma$ is the 
amplitude of the error and $\varphi_{nm}$ are the correct phases.
\begin{figure}
\epsfig{file=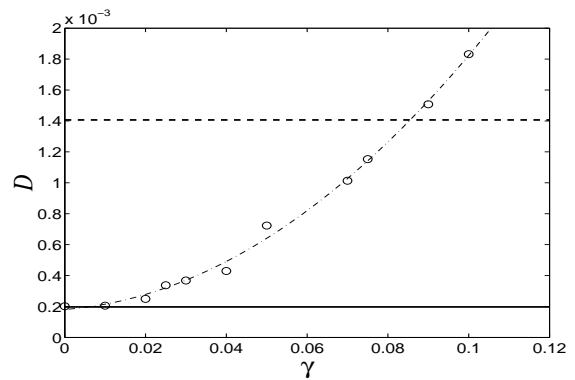,width=7.5cm,height=5.0cm}
\caption{Diffusion coefficient $D$ versus the phase error $\gamma$
in the expression (\ref{f_2}) of $f_2$ for $a = 0.7$. 
The dashed line corresponds to the case without control term, the solid line 
corresponds to the case with the full control term and the dash-dotted line
corresponds to a quadratic interpolation.}
\label{D_phase_err}
\end{figure}
Figure \ \ref{D_phase_err} shows the diffusion coefficient
versus the phase error $\gamma$ for a fixed value of $a$~($a=0.7$). 
We notice that the chaotic diffusion is still significantly 
reduced by the control
with a small error on the phases. The diffusion coefficient
is still strongly reduced by a factor greater than $2$ 
for a phase error of $5\%$. For small values of $\gamma$ the
diffusion coefficient versus $\gamma$ is well fitted by a 
quadratic interpolation, that is $D(\gamma)=D_0+D_1\gamma^2$. 
\section{conclusions}
\begin{figure}
\epsfig{file=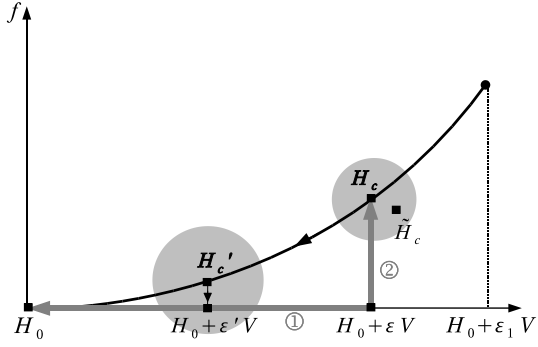,width=7.5cm,height=5.cm}
\caption{Global picture of the control: The bold curved  segment of 
curve represents a set of integrable Hamiltonians around $H_0$. 
The gray circles are the domains of existence of a given invariant torus 
around an integrable Hamiltonian. The gray arrows represent two ways of 
controlling the Hamiltonian $H_0+\varepsilon V$: the first one of order 
$\varepsilon$ and the second one of order $\varepsilon^2$.
The  arrows from $H_c$ to $H^{\prime}_c$ and from $H^{\prime}_c$
to $H_0+\varepsilon^{\prime} V$
represent  close to identity canonical transformations.}
\label{fig:concl}
\end{figure}

We have provided an effective new strategy to
control the chaotic diffusion in Hamiltonian dynamics using  small
perturbations. Since the formula
of the control term is explicit, we are able to compare the dynamics
without and with control. 
The idea of the control is pictorially represented 
in Fig.~\ref{fig:concl}: A Hamiltonian $H_0+\varepsilon V$ is controlled 
by adding a control term $f$. The naive choice for a control term would be
$f=-\varepsilon V$ but this would be useless since it is of the same 
magnitude of the source of chaotic transport and thus 
would require a major modification of the physical condition of the system of interest. In this article, we have presented a way to design an integrable controlled Hamiltonian $H_c$ with a {\em small} control term $f$ of order $\varepsilon^2$. This controlled Hamiltonian is conjugate to $H_0$ (we assume for simplicity that ${\mathcal R}V=0$). This construction of the controlled Hamiltonian works 
well up to some value $\varepsilon_1$. 
Moreover, we have shown that the control is robust, in the sense that one can use an
approximate controlled Hamiltonian $\tilde{H}_c$ which is not integrable 
but --being sufficiently close to $H_c$ -- generates a more regular dynamics (presence of invariant tori) with respect to $H_0+\varepsilon V$.
For instance, we have shown that one can successfully
use a truncated control term of 
order $\epsilon^2$ and  that one is allowed to tailor it to
some specific requirements on its shape, 
on the energy necessary to achieve control and also according to a 
partial 
knowledge of $V$(e.g. a truncation of the Fourier series and an error
on the phases).\\
The invariant tori that have been created by adding the control term 
$f$ of order $\varepsilon^2$ are those which were broken by 
increasing the amplitude of the 
perturbation, meaning that these tori are those of a Hamiltonian 
$H_0+\varepsilon^{\prime} V$ 
where $\varepsilon^{\prime} < \varepsilon$ (up to some smooth 
canonical transform close to the identity transform). 
In order to illustrate this statement, we have plotted in gray two regions of 
existence of a given invariant torus (specified e.g. by its frequency). 
The uncontrolled Hamiltonian $H_0+\varepsilon V$ does not have this invariant 
torus whereas the controlled one 
$H_c=H_0+\varepsilon V +\varepsilon^2 f$ does. 
The controlled Hamiltonian $H_c$ is conjugate to the controlled
Hamiltonian $H^{\prime}_c=H_0+\varepsilon^{\prime}V+\varepsilon^{\prime 2}f$ 
for $\varepsilon^{\prime}<\varepsilon$ 
(since they are both conjugate to $H_0$). 
Since $H_0+\varepsilon^{\prime} V$ is inside the ball around the integrable 
Hamiltonian $H_c^{\prime}$, the invariant torus 
of $H_0+\varepsilon^{\prime} V$ of 
the selected frequency is a small deformation of the torus of the controlled 
Hamiltonian $H_c^{\prime}$ and hence a small deformation of the torus
of the controlled Hamiltonian $H_c$. 
Therefore the invariant tori of $H_c$ obtained by means the control are small 
deformations of the tori of $H_0+\varepsilon V$ which were broken 
by increasing $\varepsilon$.\\
\indent We have applied this general technique of control to a 
 specific model, describing anomalous
electric transport in magnetized plasmas. In particular, we have shown
that the control term is robust, meaning that one is able to simplify it,
to reduce its amplitude or to make a small error without changing its overall
action of reducing chaotic transport.
Even though we use a
rather simplified model to describe chaotic transport of charged
particles in fusion plasmas, our result makes us believe that 
through some small smart modification of the electric potential 
 a relevant reduction of
the turbulent losses of energy and particles in tokamaks could be attained,
for the moment at least in principle.
\begin{acknowledgments}
The work reported in this article is an ongoing collaboration between
the University of Florence, the Arcetri Astrophysical Observatory (INAF),
the Centre for Theoretical Physics (Marseille) and
the Department of Research on the Controlled Fusion (CEA Cadarache).\\ 
\indent We acknowledge the financial support from Euratom/CEA 
(contract $V3382.001$),
from the italian I.N.F.N. and I.N.F.M. 
G.C. thanks I.N.F.M. for financial support through a PhD fellowship.
\end{acknowledgments}
\appendix
\section{Proof of proposition~\ref{proposition3} and \ref{proposition4}}

\subsection{Proof of proposition~\ref{proposition3}}

The terms of the series (\ref{exp_f_rv_0}) 
can be written in the following form:
$$
f_s=\sum_{n,m,k} f_{n m k}^{(s)} \sin\left[2\pi(n x+my)+\varphi_{nmk}^{(s)}-2\pi kt\right],
$$
where the sum over $k$ is from $-s$ to $s$, and the two sums over $n$ and $m$ are 
from $-sN$ to $sN$. From the recursion formula (\ref{recursion}), we have 
\begin{eqnarray}
\label{eqnapp1}
f_{n' m' k'}^{(s)}&=&\frac{a}{2s}\sum_{n,m=1}^N \frac{mn'-nm'}{(n^2+m^2)^{3/2}}
\nonumber\\
&&\times \left(-f_{n'-n,m'-m,k'-1}^{(s-1)}+f_{n-n',m-m',k'+1}^{(s-1)}\right).\nonumber\\
\end{eqnarray}
We use the following norm~:
$$
\Vert f_s\Vert=\sup_{n,m,k}\vert f_{nmk}^{(s)}\vert.
$$
From Eq.~(\ref{eqnapp1}), we get
$$
\Vert f_s\Vert \leq \frac{a \Vert f_{s-1}\Vert}{s}\sup_{n',m'}\sum_{n,m=1}^N 
\frac{\vert mn'-nm'\vert }{(n^2+m^2)^{3/2}}.
$$
Since, $\vert mn'-nm'\vert \leq sN(m+n)$, we have
$$
\Vert f_s\Vert \leq a \lambda \Vert f_{s-1}\Vert,
$$
where 
\bq
\label{eqnapp2}
\lambda=2N\sum_{n,m=1}^N \frac{m}{(n^2+m^2)^{3/2}}.
\eq
It follows that
$$
\Vert f_s\Vert \leq (a \lambda)^{s-1} a 2^{-3/2},
$$
since $\Vert f_1\Vert =a2^{-3/2}$.
Therefore, the series~(\ref{exp_f_rv_0}) converges for $a< 1/\lambda$. We notice that for $N=25$, $\lambda\approx 135$, i.e.\ the series converges for
$a\lesssim 7\times 10^{-3}$.\\
Since $n\mapsto  m/(n^2+m^2)^{3/2}$ is a positive and decreasing function, we have
$$
\sum_{n=1}^N \frac{m}{(n^2+m^2)^{3/2}}\leq\int_0^N \frac{m dn}{(n^2+m^2)^{3/2}} \leq \int_0^{\infty} \frac{m dn}{(n^2+m^2)^{3/2}}.
$$
By rescaling the integral ($t=n/m$) and using the fact that $\int_0^{\infty} \frac{dt}{(t^2+1)^{3/2}}=1$,
we have
\bq
\lambda\leq 2N\sum_{m=1}^N \frac{1}{m}\leq 2N\log N +2\gamma N +1,
\eq
where $\gamma$ is the Euler-Mascheroni constant. In particular, we notice that the bound on $\lambda$ increases like $N\log N$.\\

\subsection{Proof of proposition~\ref{proposition4}}
Concerning the regularity of the functions $f_s$ and $f$, we would like 
to show that
\bq
\label{eqnapp3}
\vert f_{nmk}^{(s)}\vert \leq \frac{a^s C^s }{(n^2+m^2)^{3/2}}.
\eq
We notice that Eq.~(\ref{eqnapp3}) is satisfied for $f_1=V$ given by 
Eq.~(\ref{potential}) for $C\geq 1$.
The inequality~(\ref{eqnapp1}) gives
\begin{eqnarray*}
&& \vert f_{n'm'k'}^{(s)}\vert \leq a^s \frac{C^{s-1}}{s} \\
&&\times \sum_{n,m=1}^N 
\frac{\vert mn'-nm'\vert }{(n^2+m^2)^{3/2}[(n'-n)^2+(m'-m)^2]^{3/2}}.
\end{eqnarray*}
We can always assume that both $n'$ and $m'$ are positive since we have 
$\vert n'-n\vert\geq \vert \vert n'\vert -n\vert$ which gives
\begin{eqnarray*}
&& \vert f_{n'm'k'}^{(s)}\vert \leq N a^s C^{s-1} \\
&&\times \sum_{n,m=1}^N 
\frac{m+n }{(n^2+m^2)^{3/2}
[(\vert n'\vert -n)^2+(\vert m'\vert -m)^2]^{3/2}},
\end{eqnarray*}
where we have used the inequality $\vert mn'-nm'\vert \leq sN(m+n)$.\\
We have to distinguish the following cases: $(i)$ $n'> N$ and $m'>N$, $(ii)$ $n'\leq N$ and $m'>N$, $(iii)$ $n'\leq N$ and $m'\leq N$. We notice that by symmetry the case $n'>N$ and $m'\leq N$ is similar to $(ii)$.\\ 
For $n',m'> N$, we have $n'\leq (N+1)(n'-n)$ and $m'\leq (N+1)(m'-m)$. Thus we have
$$
\frac{1}{(n'-n)^2+(m'-m)^2}\leq
 \frac{(N+1)^2}{n^{\prime 2}+m^{\prime 2}},
$$
which leads to
$$
\vert f_{n'm'k'}^{(s)} \vert \leq \frac{a^s C^{s-1} (N+1)^3\lambda }{(n^{\prime 2}+m^{\prime 2})^{3/2} },
$$
where $\lambda$ is given by Eq.~(\ref{eqnapp2}).\\
For $n'\leq N$ and $m'>N$, we have $m'\leq (N+1)(m'-m)$.
By using the estimate $(n'-n)^2+(m'-m)^2\geq (m'-m)^2$, we have 
$$
\frac{1}{(n'-n)^2+(m'-m)^2}\leq \frac{(N+1)^2}{m^{\prime 2}},
$$
and since $n'\leq m'$,
$$
\frac{(N+1)^2}{m^{\prime 2}}\leq \frac{2(N+1)^2}{n^{\prime 2}+m^{\prime 2}}.
$$
Thus we have 
$$
\vert f_{n'm'k'}^{(s)} \vert \leq \frac{a^s C^{s-1} 2^{3/2}(N+1)^3\lambda }{(n^{\prime 2}+m^{\prime 2})^{3/2} }.
$$
For $n'\leq N$ and $m'\leq N$, we use the crude estimates 
$$
\frac{1}{(n'-n)^2+(m'-m)^2}\leq 1,
$$
and
$$
1\leq \frac{2N^2}{n^{\prime 2}+m^{\prime 2}}.
$$
Thus we have 
$$
\vert f_{n'm'k'}^{(s)} \vert \leq \frac{a^s C^{s-1} 2^{3/2}N^3\lambda }{(n^{\prime 2}+m^{\prime 2})^{3/2} }.
$$
By denoting $C=2^{3/2}(N+1)^3\lambda$, Eq.~(\ref{eqnapp3}) is satisfied for all $n$, $m$ and for all $s$.\\ 
It follows that for $a<1/C$, the same inequality holds for $f$:
$$
\vert f_{nmk} \vert \leq \frac{C_{\infty}}{(n^2+m^2)^{3/2} },
$$
where $C_\infty=a^2C^2/(1-aC)$.


\begin{thebibliography}{99}
\bibitem{limapet} R. Lima and M. Pettini, Phys. Rev. {\bf A41}, 726
(1990); Int. J. Bif. \& Chaos {\bf 8}, 1675 (1998); 
L. Fronzoni, M. Giocondo, M. Pettini, Phys. Rev. A {\bf 43}, 6483 (1991).
\bibitem{review} G. Chen and X. Dong, Int. J. Bif. \& Chaos {\bf 3},
1363 (1993).

\bibitem{caha} J.R. Cary, Phys. Rev. Lett. {\bf 49}, 276 (1982); J.D. Hanson and J.R. Cary, Phys. Fluids {\bf 27}, 767 (1984); J.R. Cary and J.D. Hanson, Phys. Fluids {\bf 29}, 2464 (1986); C.C. Chow and J.R. Cary, Phys. Rev. Lett. {\bf 72}, 1196 (1994); W. Wan and J.R. Cary, Phys. Rev. Lett. {\bf 81}, 3655 (1998); S.R. Hudson, D.A. Monticello, A.H. Reiman, A.H. Boozer, D.J. Strickler and M.C. Zarnstorff, Phys. Rev. Lett. {\bf 89}, 275003 (2002).

\bibitem{upo} Y.C. Lai, M. Ding and C. Grebogi, Phys. Rev. E {\bf 47}, 86 (1993); Y.C. Lai, T. T\'el and C. Grebogi, Phys. Rev. E {\bf 48}, 709 (1993); O.J. Kwon, Phys. Lett. A {\bf 258}, 229 (1999); Y. Zhang and Y. Yao, Phys. Rev. E {\bf 61}, 7219 (2000); B. Doyon and L.J. Dub\'e, Phys. Rev. E {\bf 65}, 037202 (2002).

\bibitem{emb} Y. Zhang, S. Chen and Y. Yao, Phys. Rev. E {\bf 62}, 2135 (2000); J.H.E. Cartwright, M.O. Magnasco and O. Piro, Phys. Rev. E {\bf 65}, R045203 (2002).

\bibitem{gallavotti} G. Gallavotti, in {\it Regular and Chaotic Motions in Dynamical Systems}, edited by G. Velo and A.S. Wightman (Plenum, New York, 1985);
G. Gallavotti, Commun. Math. Phys. {\bf 87}, 365 (1982);
G. Gentile and V. Mastropietro, Rev. Math. Phys. {\bf 8}, 393 (1996).
\bibitem{modPulse} H. Xu, G. Wang and S. Chen, Phys. Rev. E {\bf 64}, 016201 (2001); J. Gong, H.J. W\"orner and P. Brumer, Phys. Rev. E (to appear), archived in \texttt{arxiv.org/quant-ph/0309215}.
\bibitem{modLoc} A. Oloumi and D. Teychenn\'e, Phys. Rev. E {\bf 60}, R6279 (1999).
\bibitem{opt} C.D. Schwieters and H. Rabitz, Phys. Rev. A {\bf 44}, 5224 (1991); J. Botina, H. Rabitz and N. Rahman, Phys. Rev. A {\bf 51}, 923 (1995); E.M. Bollt and J.D. Meiss, Physica D {\bf 81}, 280 (1995); E.M. Bollt and J.D. Meiss, Phys. Lett. A {\bf 204}, 373 (1995).

\bibitem{modExt} Z. Wu, Z. Zhu and C. Zhang, Phys. Rev. E {\bf 57}, 366 (1998); L. Sirko and P.M. Koch, Phys. Rev. Lett. {\bf 89}, 274101 (2002). 
\bibitem{marc88} M. Pettini, A. Vulpiani, J.H. Misguich, M. De Leener, J. Orban, and R. Balescu, Phys. Rev. A {\bf 38}, 344 (1988); 
M. Pettini, in {\em Non-Linear
Dynamics}, edited by G. Turchetti, (World Scientific,
Singapore, 1989).
\bibitem{scott} B.D.\ Scott, Phys. Plasmas {\bf 10}, 963 (2003) and the list of references $[3]$ therein.
\bibitem{ws_chaos}A.J. Lichtenberg and M.A. Lieberman, {\it Regular and Stochastic motion} (Springer-Verlag, Berlin, 1993).

\bibitem{Vulpiani} A. Crisanti, M. Falcioni, G. Paladin and A. Vulpiani, 
La Rivista del Nuovo Cimento {\bf 14}, 1 (1991);  
 T. Bohr, M. H. Jensen, G. Paladin, A. Vulpiani, {\it Dynamical Systems Approach to Turbulence} (Cambridge University Press, Cambridge, 1998); 
M. H\'enon, CR Acad. Sci. Paris A {\bf 262}, 312 (1966).

\bibitem{exp_chaos} R. McWilliams and M. Okubo, Phys. Fluids {\bf 30}, 2849 (1987).

\bibitem{philippe1} Ph. Ghendrih, Y. Sarazin, G. Attuel, S. Benkadda, P. Beyer, G. Falchetto, C. Figarella, X. Garbet, V. Grandgirard and M. Ottaviani, Nuclear Fusion, {\bf 43}, 1013 (2003).
\bibitem{philippe2} JET team, Nuclear Fusion, {\bf 39}, 1891 (1999).

\bibitem{Northrop} T.J. Northrop, Annals of Physics {\bf 15}, 79 (1961).

\bibitem{Amato} E. Amato, M. Pettini and M. Salvati, Astron. Astrophys. 
{\bf 402}, 819 (2003). 

\bibitem{michel} M. Vittot, {\it Perturbation Theory and Control in
Classical or Quantum Mechanics by an Inversion Formula}, archived in
\texttt{arXiv.org/math-ph/0303051}

\bibitem{anormal_exp} A.J. Wootton, H. Matsumoto, K. McGuire, W.A. Peebles, Ch.P. Ritz, P.W. Terry and S.J. Zweben, Phys. Fluids B {\bf 2}, 2879 (1990).

\bibitem{guzz02}
 M. Guzzo, E. Lega, and C. Froeschl\'e, Physica D {\bf 163}, 1 (2002).
\bibitem{guido1} G.Ciraolo, C. Chandre, R. Lima, M. Vittot and M. Pettini, {\em Control of chaos in Hamiltonian systems},  archived in \texttt{arXiv.org/nlin.CD/0311009}.
\bibitem{guido2} G.Ciraolo, C. Chandre, R. Lima, M. Vittot, M. Pettini, Charles Figarella, and Philippe Ghendrih {\em Control of chaotic transport in Hamiltonian systems},  archived in \texttt{arXiv.org/nlin.CD/0304040}.

\end{thebibliography}
\end{document}